\documentclass[conference,9pt]{IEEEtran}
\IEEEoverridecommandlockouts

\usepackage{cite}
\usepackage{amsmath,amssymb,amsfonts}
\usepackage{algorithm}
\usepackage{graphicx}
\usepackage{textcomp}
\usepackage{xcolor}

\usepackage{booktabs}
\usepackage[hidelinks]{hyperref}
\usepackage{siunitx}
\usepackage{mathtools}
\usepackage{comment}
\usepackage{multirow}
\usepackage{makecell}
\usepackage{flushend}

\usepackage{color}
\usepackage{multirow}
\usepackage{algpseudocode}
\usepackage{url}
\usepackage{amssymb,amsmath,bm}
\usepackage{textcomp}
\usepackage{cite}
\usepackage{CJKutf8}
\usepackage{hhline}
\usepackage{bbding}
\usepackage{pifont}
\usepackage{wasysym}
\usepackage{arydshln}

\makeatletter
\let\MYcaption\@makecaption
\makeatother
\usepackage[font=footnotesize,subrefformat=parens]{subcaption}
\makeatletter
\let\@makecaption\MYcaption
\makeatother

\usepackage{xspace}
\makeatletter
\DeclareRobustCommand\onedot{\futurelet\@let@token\@onedot}
\def\@onedot{\ifx\@let@token.\else.\null\fi\xspace}

\makeatother

\def\equationautorefname~#1\null{(#1\null)}

\renewcommand{\sectionautorefname}{Section}
\renewcommand{\subsectionautorefname}{\sectionautorefname}

\def\L{{\cal L}}

\newcommand{\maru}[1]{\raise0.2ex\hbox{\textcircled{\scriptsize{#1}}}}

\let\orgautoref\autoref

\renewcommand{\autoref}[1]
{%
\def\figureautorefname{Fig.}%
\def\subfigureautorefname{\figureautorefname}%
\def\sectionautorefname{Sec.}%
\def\subsectionautorefname{\sectionautorefname}%
\def\subsectionautorefname{\sectionautorefname}%
\orgautoref{#1}%
}

\def\appendixautorefname~#1\null{~#1 \null}

\makeatletter
\newcommand{\figcaption}[1]{\def\@captype{figure}\caption{#1}}
\newcommand{\tblcaption}[1]{\def\@captype{table}\caption{#1}}
\makeatother

\makeatletter 
\newcommand{\linebreakand}{%
  \end{@IEEEauthorhalign}
  \hfill\mbox{}\par
  \mbox{}\hfill\begin{@IEEEauthorhalign}
}
\makeatother 

\def\BibTeX{{\rm B\kern-.05em{\sc i\kern-.025em b}\kern-.08em
    T\kern-.1667em\lower.7ex\hbox{E}\kern-.125emX}}
\begin{document}

\title{All-in-One ASR: Unifying Encoder-Decoder Models of \\ CTC, Attention, and Transducer in Dual-Mode ASR}

\author{\IEEEauthorblockN{Takafumi Moriya, Masato Mimura, Tomohiro Tanaka, Hiroshi Sato, Ryo Masumura, Atsunori Ogawa}
\IEEEauthorblockA{NTT, Inc., Japan}}

\maketitle
\begin{abstract}
This paper proposes a unified framework, All-in-One ASR, that allows a single model to support multiple automatic speech recognition (ASR) paradigms, including connectionist temporal classification (CTC), attention-based encoder-decoder (AED), and Transducer, in both offline and streaming modes. While each ASR architecture offers distinct advantages and trade-offs depending on the application, maintaining separate models for each scenario incurs substantial development and deployment costs. To address this issue, we introduce a multi-mode joiner that enables seamless integration of various ASR modes within a single unified model. Experiments show that All-in-One ASR significantly reduces the total model footprint while matching or even surpassing the recognition performance of individually optimized ASR models. Furthermore, joint decoding leverages the complementary strengths of different ASR modes, yielding additional improvements in recognition accuracy.
\end{abstract}


\section{Introduction}
End-to-end automatic speech recognition (ASR) approaches have gained attention in the ASR research community~\cite{jinyu2022survey,rohit2024suvey}, leading to various models such as connectionist temporal classification (CTC)\cite{Graves2006}, attention-based encoder-decoder (AED)\cite{chorowski2014,vaswani2017att}, and Transducer~\cite{Graves2012}.
Although each ASR model has distinct advantages and limitations depending on the application, developing and maintaining them separately entails substantial overhead. 
Moreover, as model size increases, the development and maintenance overhead associated with each ASR mode explodes.
Costs are further exacerbated when separately trained models are expected to support both offline and streaming modes, particularly in resource-constrained scenarios such as on-device ASR~\cite{Sainath2021AnES}.

Dual-mode ASR~\cite{yu2021dualmode} is a solution that unifies separate offline and streaming ASR models into a single unified model; the dual-mode ASR model enables both offline and streaming processing. In offline mode, the encoder processes the entire input acoustic feature sequence. In contrast, in streaming mode, it processes the input sequentially from left to right, with only limited access to future context. The dual-mode ASR framework can halve the model size.

The joint training of multiple ASR models not only enables the sharing of partial modules but also enhances training efficiency and ASR performance by leveraging their complementary strengths~\cite{Kim2017JointCB,watanabe2017hybrid,karita2020ctctransformer,tanaka2019aje,tanaka2019joint_dnn_aed,moritz19trigger,niko2019beamalgo,niko2020streaming,tara2019twopass,Hu2020Deliberation,moriya2021rnnts2s,moriya2022rnntadlm,tang2023hybrid_rnnt_aed,moritz2023meta_rnnt,sudo2025asr4d}. 
As a representative example, AED tends to produce stronger cross-attention weights for short utterances, resulting in better ASR performance~\cite{prabhavalkar2017transducer_with_attention,tara2019twopass}, whereas the Transducer is more robust for long-form ASR since it does not rely on cross-attention~\cite{chiu2019comp,inaguma2023AlignmentKD,stooke2024aligner}. 
To leverage these attributes, some approaches perform joint decoding using Transducer and AED decoder branches with a shared encoder, leading to improved ASR performance~\cite{tara2019twopass,Hu2020Deliberation,moriya2021rnnts2s,moriya2022rnntadlm,tang2023hybrid_rnnt_aed}.
However, although joint training brings benefits to both training and decoding, it requires extra parameters for each decoder branch. 
Furthermore, each decoder branch of the autoregressive ASR models must compute decoder states separately, which inevitably results in higher computation costs, increased memory usage for caching, and slower decoding. 

In this paper, we address the challenge of unifying different ASR characteristics into a single model without using separate decoder branches, including the distinctions between autoregressive models (AED and Transducer) and non-autoregressive models (CTC), as well as between models with cross-attention (AED) and those without it (CTC and Transducer). 
To realize this, we propose an All-in-One ASR framework that unifies not only offline and streaming modes but also various ASR models, such as CTC, AED, and Transducer. 
The proposed method employs a Transducer model as its foundation architecture, and its key advance is the introduction of a multi-mode joiner. 
The multi-mode joiner emulates CTC, AED, or Transducer behavior by selecting the mode while fully sharing parameters across modes, significantly reducing the overall model size compared to building separate ASR models for each mode. 
Moreover, the All-in-One ASR model shares the predictor across the AED and Transducer modes, eliminating redundant recomputation of decoder states for identical hypotheses and enabling efficient joint decoding.

Our contributions are as follows:
\begin{itemize}
\item We propose an All-in-One ASR framework that unifies CTC, AED, and Transducer in both offline and streaming modes into a single unified model, without requiring additional parameters. 
\item We compare encoder (attention vs. attention-free) and predictor (LSTM vs. Mamba) types of interest in the ASR community. 
\item Experimental results demonstrate that our All-in-One ASR model achieves performance comparable to that of each separately built ASR model, while reducing the overall model size.
\item Joint decoding across different ASR modes within the All-in-One ASR model further improves performance by leveraging the strengths of each mode to complement each other. 
\end{itemize}


\section{Preliminaries}
\label{ssec:preliminaries}
\vspace{-0.025cm}
\subsection{Transducer-based ASR}
\label{sssec:hat}
\vspace{-0.075cm}
Our approach is grounded in the Transducer model~\cite{Graves2012}. In this study, we employ the Hybrid Autoregressive Transducer (HAT)~\cite{Ehsan2020hat}. 
First, the encoder, $f^{\text{enc}}(\cdot)$, transforms the input acoustic feature sequence $\bm{X} = \left[ \bm{x}_1, \dots, \bm{x}_{T} \right]$ of length-$T$ into $\bm{H}^{\text{enc}} = \left[ \bm{h}^{\text{enc}}_{1}, \dots, \bm{h}^{\text{enc}}_{T} \right] \in \mathbb{R}^{T \times D^{\prime}}$. 
Then, the predictor, $f^{\text{pred}}(\cdot)$, embeds non-blank (label) token sequence $Y = \left[ y_1, \dots, y_U \right]$ of length-$U$, where $y_u \in \mathcal{Y}$, into $\bm{H}^{\text{pred}} = \left[ \bm{h}^{\text{pred}}_{1}, \dots, \bm{h}^{\text{pred}}_{U} \right] \in \mathbb{R}^{U \times D^{\prime \prime}}$. 
Here, $\mathcal{Y}$ is the token set that consists of $K$ tokens, excluding the blank symbol $\phi$. 
$D^{\prime}$ and $D^{\prime \prime}$ indicate the output dimensions of each network.
The joiner produces the prediction $\hat{\bm{y}}_{t,u} \in (0,1)^{K+1}$ based on the $t$-th acoustic and $u$-th linguistic encoded features, as follows:
\begin{eqnarray}
\bm{h}^{\text{enc}}_{t} &=& f^{\text{enc}} (\bm{x}_{t}; \theta^{\text{enc}}), \label{eq:enc} \\
\bm{h}^{\text{pred}}_{u} &=& f^{\text{pred}} (y_{u-1}; \theta^{\text{pred}}), \label{eq:pred} \\
\hat{\bm{y}}_{t,u} &=& f^{\text{joiner}} (\bm{h}^{\text{enc}}_{t}, \bm{h}^{\text{pred}}_{u}; \theta^{\text{joiner}}), \label{eq:joiner}
\end{eqnarray}
where $\theta^{\text{enc}}$, $\theta^{\text{pred}}$, and $\theta^{\text{joiner}}$ indicate learnable parameters. 
In the training step, the predictions form three dimensional tensor $\hat{\bm{Y}}^{\text{HAT}} \in (0,1)^{T \times U \times (K+1)}$. 
The learnable parameter set $\Theta^{\text{HAT}} \triangleq \{ \theta^{\text{enc}}, \theta^{\text{pred}}, \theta^{\text{joiner}} \}$ is optimized using Transducer loss $\mathcal{L}_{\text{HAT}}$~\cite{Graves2012}. 

In this work, the HAT model serves as the foundation architecture, and its joiner plays an important role in the proposed framework.
We describe the specific processing of the standard joiner below.
The HAT model separately outputs blank probability $\hat{y}_{t,u}^{\text{blank}} \in (0,1)^{1}$ and label probabilities $\hat{\bm{y}}_{t,u}^{\text{label}} \in (0,1)^{K}$, as follows:
\begin{eqnarray}
\bm{h}_{t,u}^{\text{joiner}} &=& \text{tanh} (\bm{W}^{\text{enc}} \bm{h}^{\text{enc}}_{t} + \bm{W}^{\text{pred}} \bm{h}^{\text{pred}}_{u} + \bm{b}^{\text{enc}} + \bm{b}^{\text{pred}}), \label{eq:joiner_space_hat} \\
\hat{y}_{t,u}^{\text{blank}} &=& \text{Sigmoid} (\bm{w}^{\text{blank}} \bullet \bm{h}_{t,u}^{\text{joiner}} + b^{\text{blank}} ), \label{eq:blank} \\ 
\hat{\bm{y}}_{t,u}^{\text{label}} &=& \text{Softmax} (\bm{W}^{\text{label}} \bm{h}_{t,u}^{\text{joiner}} + \bm{b}^{\text{label}}), \label{eq:nonblank} \\
\hat{\bm{y}}_{t,u} &=& \left[ \ \hat{y}_{t,u}^{\text{blank}}, \ (1 - \hat{y}_{t,u}^{\text{blank}}) \cdot \hat{\bm{y}}_{t,u}^{\text{label} \ \top} \ \right]^{\top}, \label{eq:concat} 
\label{eq:hat_mode}
\end{eqnarray}
where $\{ \bm{W}^{\text{enc}} \in \mathbb{R}^{D \times D^{\prime}}, \bm{W}^{\text{pred}} \in \mathbb{R}^{D \times D^{\prime \prime}}, \bm{b}^{\text{enc}} \in \mathbb{R}^{D}, \bm{b}^{\text{pred}} \in \mathbb{R}^{D} \}$ is the parameter set of the joiner space, which produces $\bm{h}_{t,u}^{\text{joiner}}$ via the hyperbolic tangent function $\text{tanh}(\cdot)$. 
$\hat{y}_{t,u}^{\text{blank}}$ is computed using weight vector $\bm{w}^{\text{blank}} \in \mathbb{R}^{D}$ and bias term $b^{\text{blank}} \in \mathbb{R}^{1}$, where $\bullet$ and $\text{Sigmoid}(\cdot)$ denote the dot-product and sigmoid operation, respectively. 
$\{ \bm{W}^{\text{label}} \in \mathbb{R}^{K \times D}, \bm{b}^{\text{label}} \in \mathbb{R}^{K} \}$ are the parameters of the linear output layer, which yields $\hat{\bm{y}}_{t,u}^{\text{label}}$ via the softmax function $\text{Softmax}(\cdot)$.
Thus, the standard joiner parameter set $\theta^{\text{joiner}}$ is defined as $\theta^{\text{joiner}} \triangleq \{ \bm{W}^{\text{enc}}, \bm{W}^{\text{pred}}, \bm{b}^{\text{enc}}, \bm{b}^{\text{pred}}, \bm{w}^{\text{blank}}, b^{\text{blank}}, \bm{W}^{\text{label}}, \bm{b}^{\text{label}} \}$.

\begin{figure}[t]
    \centering
    \includegraphics[width=9.0cm]{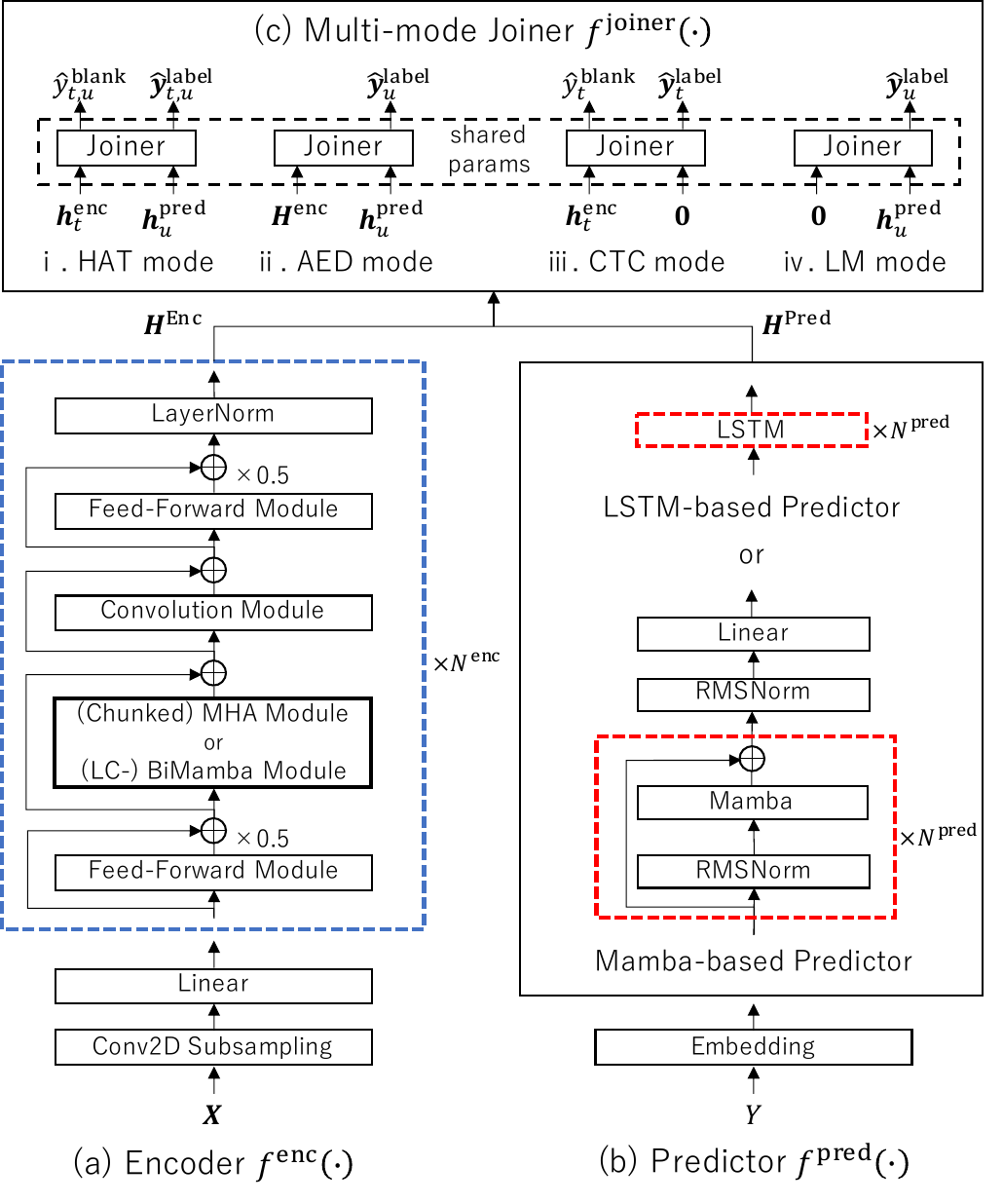}
\vspace{-0.6cm}
 \caption{Overview of the foundational Transducer architecture, consisting of (a) Encoder, (b) Predictor, and (c) Joiner. The proposed multi-mode joiner supports HAT, AED, CTC, and LM behaviors by switching modes.}
 \label{fig:architecture}
\end{figure}

\subsection{Dual-mode ASR}
\label{sssec:dualmode}
Dual-mode ASR~\cite{yu2021dualmode} is a key framework that enables the unification of the parameter sets for offline and streaming ASR into a single shared parameter set.
For dual-mode ASR training, the unified parameter set is optimized using the combined loss $\L = \L^{\text{Offline}} + \L^{\text{Streaming}}$, where $\L^{\text{Offline}}$ and $\L^{\text{Streaming}}$ represent the offline and the streaming mode losses, respectively. 
To compute the loss for each mode, encoder outputs and corresponding predictions must be independently prepared for both the offline and streaming modes.
Thus, the switching between offline and streaming modes depends solely on encoder processing, as detailed below.

\subsection{Details of encoder processing for offline and streaming modes}
\label{ssec:encoder}
In this work, we adopt \textit{self-attention-based Conformer}~\cite{anmol2020conformer} and \textit{attention-free ConMamba}~\cite{jiang2025speechslytherin_conmamba,zhang2024conmamba} encoders, as illustrated in Fig.~\ref{fig:architecture} (a). The key difference is that the multi-headed self-attention (MHA) module within the Conformer block is replaced with the Mamba module~\cite{mamba}, a selective state-space model. The offline Conformer and ConMamba encoders straightforwardly employ full self-attention and bidirectional Mamba (BiMamba), respectively. Note that the batch normalization~\cite{ioffe2015batchnorm} within the convolution module is replaced with layer normalization~\cite{jimmy2016layernorm}. Hence, in this section, we briefly describe the chunkwise encoder processing for the streaming mode. 

For streaming ASR, both encoders operate in a chunkwise manner to enable real-time processing. 
The chunkwise Conformer is implemented by using an attention mask~\cite{chen2021lcconformer,yangyang2022emformer,strimel2023lookahead,qian2020streaming_tt,Li2023DynamicConv,shashi2025dualmode_xlsr,mimura2025attn_mamba} during training. This mask drops the left and right contexts beyond a predefined length. 
To adapt BiMamba for streaming ASR, chunking is applied to the input features. We refer to this variant as latency-controlled BiMamba (LC-BiMamba)~\cite{moriya2025lcssm_dual}; our inspiration is drawn from long short-term memory (LSTM)~\cite{hochreiter1997lstm} with chunkwise processing~\cite{yu2016lcblstm,Xue2017lcblstm}. 
Note that in streaming mode, the depthwise convolution within the convolution module is replaced with a causal one using identical parameters~\cite{li2021causal_online_model}.
Since both streaming Conformer and ConMamba perform chunkwise processing, the input sequence $\bm{X}$ is segmented into $I$ chunks of length $L_{c}$, denoted as $(\bm{X}_{1}, \dots, \bm{X}_{I})$, where $\bm{X}_{i} = \left[ \bm{x}_{(i-1) \times L_{c} + 1}, \dots, \bm{x}_{i \times L_{c}} \right]^{\top}$ represents the $i$-th input chunk.

\begin{figure}[t]
 \begin{center}
    \includegraphics[width=6.2cm]{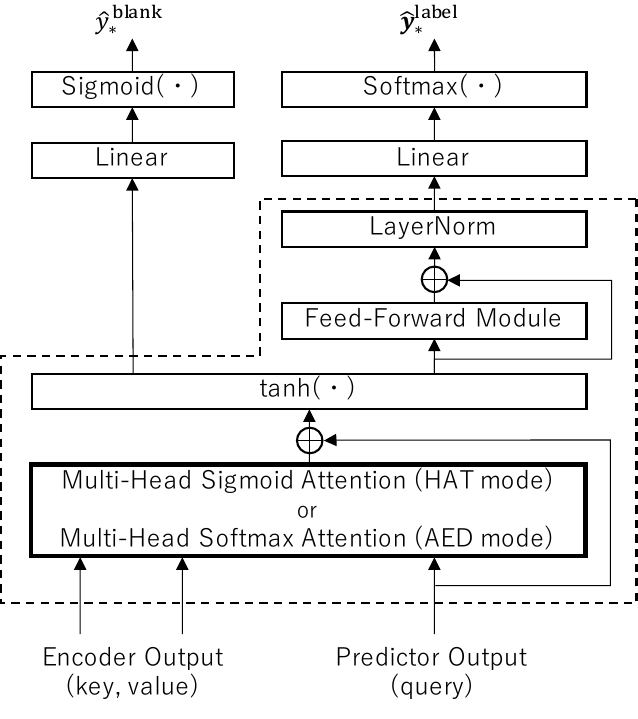}
 \end{center}
\vspace{-0.4cm}
 \caption{Architecture of the proposed multi-mode joiner. By replacing the predictor and $\text{tanh}(\cdot)$ in the dotted-line block with masked self-attention and $\text{LayerNorm}(\cdot)$, the resulting structure is equivalent to a single-layer Transformer decoder.
 }
 \label{fig:advanced_joiner}
\end{figure}

\section{Proposed multi-mode joiner for All-in-One ASR}
\label{sec:proposed}
Joint training of multiple ASR models using shared parameters is beneficial not only during training but also at the inference stage, where joint decoding can lead to improved recognition performance~\cite{Kim2017JointCB,watanabe2017hybrid,karita2020ctctransformer,tara2019twopass,tanaka2019aje,tanaka2019joint_dnn_aed,Hu2020Deliberation,moritz19trigger,niko2019beamalgo,niko2020streaming,moriya2021rnnts2s,moriya2022rnntadlm,moritz2023meta_rnnt,tang2023hybrid_rnnt_aed,sudo2025asr4d}. 
However, while the encoder can be shared, each ASR model typically requires its own decoder, resulting in an increased number of parameters. 
Moreover, for joint decoding with autoregressive models (HAT and AED), each decoder must compute its own predictions individually, resulting in slower inference. 

To address these issues, we propose an All-in-One ASR framework that incorporates a multi-mode joiner capable of operating as a Transducer, AED, CTC, or language model (LM) within a single unified model, simply by specifying the mode, as illustrated in Fig.~\ref{fig:architecture} (c) i–iv.
The proposed multi-mode joiner, inspired by the Transformer-based decoder architecture~\cite{vaswani2017att} and illustrated in Fig.~\ref{fig:advanced_joiner}, enables the use of a multi-headed cross-attention mechanism.
In the following, we provide detailed descriptions of the processing for each mode in the multi-mode joiner.

\subsection{Transducer mode (HAT mode)}
\label{ssec:hat_mode}

The multi-mode joiner supports multi-headed attention for cross-attention computation, but it must avoid accessing full contexts to enable frame-synchronous decoding, particularly in streaming ASR.
To address this issue, for the HAT mode, we introduce a \textit{sigmoid cross-attention} inspired by the multiplicative integration joiner~\cite{saon2021mijoiner,yuhuai2016mulitiplicative_integration} and sigmoid self-attention~\cite{ramapuram2025sigmoid_attention}.
This extension enables framewise computation of attention weights $\alpha_{t,u}^{* \text{(HAT)}}$ and context vectors $\bm{c}_{t,u}^{* \text{(HAT)}}$ in the HAT mode. 
We reformulate and extend the joiner space for the multi-mode joiner in the HAT mode as follows:
\begin{eqnarray}
\bm{h}^{\text{enc}^{\prime}}_{t} &=& \bm{W}^{\text{enc}} \bm{h}^{\text{enc}}_{t} + \bm{b}^{\text{enc}} \in \mathbb{R}^{D}, \label{eq:advanced_enc_out}  \\
\bm{h}^{\text{pred}^{\prime}}_{u} &=& \bm{W}^{\text{pred}} \bm{h}^{\text{pred}}_{u} + \bm{b}^{\text{pred}} \in \mathbb{R}^{D}, \label{eq:advanced_pred_out} \\
\bm{q}_{u} &=& \bm{W}^{\text{query}} \ \text{LayerNorm} (\bm{h}^{\text{pred}^{\prime}}_{u} ; \theta^{\text{LN}_{\text{Q}}}) \in \mathbb{R}^{D},  \label{eq:advanced_query} \\
\bm{k}_{t} &=& \bm{W}^{\text{key}} \ \text{LayerNorm} (\bm{h}^{\text{enc}^{\prime}}_{t} ; \theta^{\text{LN}_{\text{KV}}}) \in \mathbb{R}^{D}, \label{eq:advanced_key} \\
\bm{v}_{t} &=& \bm{W}^{\text{value}} \ \text{LayerNorm} (\bm{h}^{\text{enc}^{\prime}}_{t} ; \theta^{\text{LN}_{\text{KV}}}) \in \mathbb{R}^{D}, \label{eq:advanced_value} \\
\alpha_{t,u}^{\text{Head}_h \text{(HAT)}} &=& \text{Sigmoid} \left(\frac{\bm{k}_{t}^{\text{Head}_h} \bullet \bm{q}_{u}^{\text{Head}_h}}{\sqrt{d}}\right) \in (0, 1)^{1}, \label{eq:sigmoid_attention} \\
\bm{c}_{t,u}^{\text{Head}_h \text{(HAT)}} &=& \alpha_{t,u}^{\text{Head}_h \text{(HAT)}} \cdot \bm{v}_{t}^{\text{Head}_h} \in \mathbb{R}^{d}, \\
\bm{c}_{t,u}^{\text{(HAT)}} &=& \bm{W}^{\text{proj}} \left[ \bm{c}_{t,u}^{\text{Head}_{1} \text{(HAT)} \top}, \cdots, \bm{c}_{t,u}^{\text{Head}_{d^{h}} \text{(HAT)} \top}\right]^{\top} \in \mathbb{R}^{D}, \nonumber \\
\\
\bm{h}_{t,u}^{\text{joiner (HAT)}} &=& \text{tanh} ( \bm{h}^{\text{pred}^{\prime}}_{u} + \bm{c}_{t,u}^{\text{(HAT)}}) ,
\end{eqnarray}
where $\text{LayerNorm} (\cdot)$ denotes layer normalization with learnable parameter set $\theta^{\text{LN}_{*}}$, where parameter set $\theta^{\text{LN}_{\text{KV}}}$ is shared for computing both the query and key vectors.
Each query, key, and value vector is then split into $d^{h}$ vectors of dimension $d$, where $d^{h}$ is the total number of heads and $d = D / d^{h}$. 
A framewise attention weight for each head is computed using Eq.~(\ref{eq:sigmoid_attention}) and then multiplied by the corresponding head value vector.
The resulting context vectors from all heads are concatenated and projected to context vector $\bm{c}_{t,u}^{\text{(HAT)}}$.
We then obtain $\bm{h}_{t,u}^{\text{joiner (HAT)}}$ from $\bm{h}^{\text{pred}^{\prime}}_{u}$ and $\bm{c}_{t,u}^{\text{(HAT)}}$. 
The additional parameters for the multi-headed sigmoid attention module shown in Fig.~\ref{fig:advanced_joiner} are defined as  $\theta^{\text{Attention}} \triangleq \{ \bm{W}^{\text{query}} \in \mathbb{R}^{D \times D}, \bm{W}^{\text{key}} \in \mathbb{R}^{D \times D}, \bm{W}^{\text{value}} \in \mathbb{R}^{D \times D}, \bm{W}^{\text{proj}} \in \mathbb{R}^{D \times D} \} \cup \theta^{\text{LN}_{\text{Q}}} \cup \theta^{\text{LN}_{\text{KV}}}$.

The blank probability $\hat{y}_{t,u}^{\text{blank \ (HAT)}}$ is obtained from $\bm{h}_{t,u}^{\text{joiner (HAT)}}$. 
The label probabilities $\hat{\bm{y}}_{t,u}^{\text{label \ (HAT)}}$ are computed by passing $\bm{h}_{t,u}^{\text{joiner (HAT)}}$ through a feed-forward module, $\text{FeedForward}(\cdot)$, and layer normalization, parameterized by $\theta^{\text{FF}}$ and $\theta^{\text{LN}_{\text{FF}}}$, respectively, followed by a linear output layer, as follows:
\begin{eqnarray}
\bm{h}_{t,u}^{\text{joiner}^{\prime} \text{(HAT)}} &=&  \text{FeedForward} (\bm{h}_{t,u}^{\text{joiner (HAT)}} ; \theta^{\text{FF}}), \\
\bm{h}_{t,u}^{\text{joiner}^{\prime \prime} \text{(HAT)}} &=&  \text{LayerNorm} (\bm{h}_{t,u}^{\text{joiner (HAT)}} + \bm{h}_{t,u}^{\text{joiner}^{\prime} \text{(HAT)}} ; \theta^{\text{LN}_{\text{FF}}}), \\
\hat{y}_{t,u}^{\text{blank \ (HAT)}} &=& \text{Sigmoid} (\bm{w}^{\text{blank}} \bullet \bm{h}_{t,u}^{\text{joiner (HAT)}} + b^{\text{blank}} ), \\ 
\hat{\bm{y}}_{t,u}^{\text{label \ (HAT)}} &=& \text{Softmax} (\bm{W}^{\text{label}} \bm{h}_{t,u}^{\text{joiner}^{\prime \prime} \text{(HAT)}}  + \bm{b}^{\text{label}}), \label{eq:label_prob_hat} \\
\hat{\bm{y}}^{\text{(HAT)}}_{t,u} &=& \left[ \ \hat{y}_{t,u}^{\text{blank \ (HAT)}}, \ (1 - \hat{y}_{t,u}^{\text{blank \ (HAT)}}) \cdot \hat{\bm{y}}_{t,u}^{\text{label \ (HAT)} \ \top} \ \right]^{\top}. \nonumber \\
\end{eqnarray}

Therefore, the multi-mode joiner parameter set $\theta^{\text{joiner}^{\prime}}$ is defined as $\theta^{\text{joiner}^{\prime}} \triangleq \theta^{\text{joiner}} \cup \theta^{\text{Attention}} \cup \theta^{\text{FF}} \cup \theta^{\text{LN}_{\text{FF}}}$, and the parameter set of the HAT model with the multi-mode joiner is redefined as $\Theta^{\text{HAT}^{\prime}} \triangleq \{ \theta^{\text{enc}}, \theta^{\text{pred}}, \theta^{\text{joiner}^{\prime}} \}$, which is trained using $\mathcal{L}_{\text{HAT}}$.
Note that the reason for omitting the feed-forward module and layer normalization in the computation of $\hat{y}_{t,u}^{\text{blank (HAT)}}$ is to enable extensions such as blank thresholding, which leads to faster decoding~\cite{le2023BlankThresholding,moriya2024hat_iam}. 

\subsection{AED mode}
\label{ssec:aed_mode}
To emulate AED behavior, we reformulate the joiner space using parameter set $\theta^{\text{joiner}^{\prime}}$. 
As shown in Fig.~\ref{fig:advanced_joiner}, the multi-mode joiner can be interpreted as a single-layer Transformer decoder~\cite{vaswani2017att}, assuming the predictor is regarded as a masked self-attention-based MHA module. 
The multi-headed softmax attention module for the AED mode is constructed by reusing $\theta^{\text{joiner}^{\prime}}$, which is identical to the parameters of the HAT mode, as follows:
\begin{eqnarray}
\bm{K} &=& \left[ \bm{k}_{1}, ..., \bm{k}_{T} \right]^{\top} \in \mathbb{R}^{T \times D}, \\
\bm{V} &=& \left[ \bm{v}_{1}, ..., \bm{v}_{T} \right]^{\top} \in \mathbb{R}^{T \times D}, \\
\bm{\alpha}_{u}^{\text{Head}_h \text{(AED)}} &=& \text{Softmax} \left(\frac{\bm{K}^{\text{Head}_h} \bm{q}_{u}^{\text{Head}_h}}{\sqrt{d}}\right) \in (0, 1)^{T}, \label{eq:softmax_attention} \\
\bm{c}_{u}^{\text{Head}_h \text{(AED)}} &=& \bm{V}^{\text{Head}_h \top} \bm{\alpha}_{u}^{\text{Head}_h \text{(AED)}} \in \mathbb{R}^{d}, \\
\bm{c}_{u}^{\text{(AED)}} &=& \bm{W}^{\text{proj}} \left[ \bm{c}_{u}^{\text{Head}_1 \text{(AED)} \top}, \cdots, \bm{c}_{u}^{\text{Head}_{d^{h}} \text{(AED)} \top}\right]^{\top} \in \mathbb{R}^{D}, \label{eq:context_vector_aed} \nonumber \\
\\
\bm{h}_{u}^{\text{joiner (AED)}} &=& \text{tanh} ( \bm{h}^{\text{pred}^{\prime}}_{u} + \bm{c}_{u}^{\text{(AED)}})  \label{eq:joiner_space_aed}, 
\end{eqnarray}
where query vector $\bm{q}_{u}$ is the same as in Eq. (\ref{eq:advanced_query}). 
$\bm{K}$ and $\bm{V}$ denote the key and value matrices, respectively, constructed by collecting the corresponding vectors from Eqs. (\ref{eq:advanced_key}) and (\ref{eq:advanced_value}) at each timestep. 
Then, we obtain the $u$-th prediction $\hat{\bm{y}}_{u}^{\text{label (AED)}} \in (0,1)^{K}$ as follows:
\begin{eqnarray}
\bm{h}_{u}^{\text{joiner}^{\prime} \text{(AED)}} &=&  \text{FeedForward} (\bm{h}_{u}^{\text{joiner  (AED)}} ; \theta^{\text{FF}}), \label{eq:ff_aed} \\
\bm{h}_{u}^{\text{joiner}^{\prime \prime} \text{(AED)}} &=&  \text{LayerNorm} (\bm{h}_{u}^{\text{joiner (AED)}} + \bm{h}_{u}^{\text{joiner}^{\prime} \text{(AED)}} ; \theta^{\text{LN}_{\text{FF}}}), \nonumber \\
\\
\hat{\bm{y}}_{u}^{\text{label \ (AED)}} &=& \text{Softmax} (\bm{W}^{\text{label}} \bm{h}_{u}^{\text{joiner}^{\prime \prime} \text{(AED)}}  + \bm{b}^{\text{label}}) \label{eq:label_prob_aed}.
\end{eqnarray}
In the AED mode, time index $t$ is omitted, so the output becomes two dimensional matrix $\hat{\bm{Y}}^{\text{AED}} \in (0,1)^{U \times K}$ during training, while parameter set $\Theta^{\text{HAT}^{\prime}}$ is optimized using the cross-entropy loss $\mathcal{L}_{\text{AED}}$. 

\begin{figure}[t]
 \begin{center}
    \includegraphics[width=8.4cm]{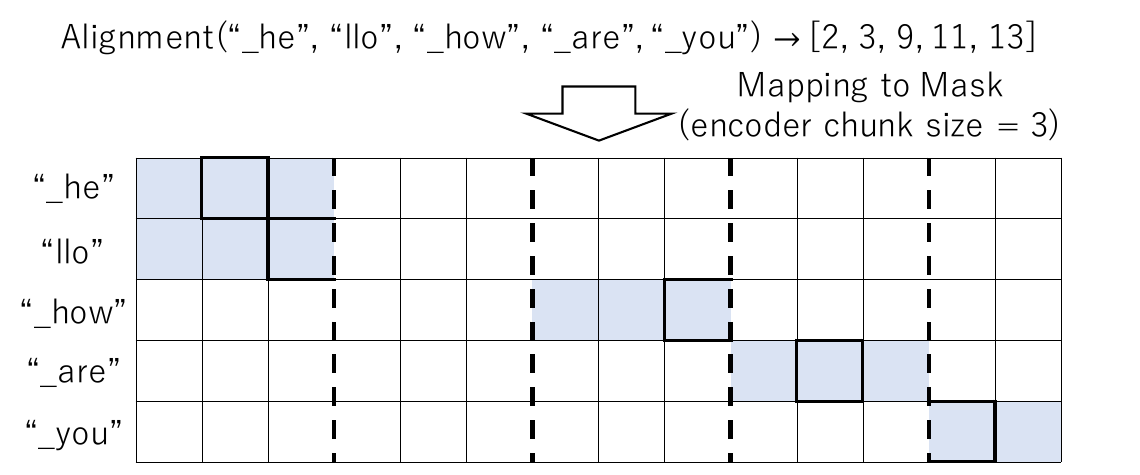}
 \end{center}
\vspace{-0.2cm}
 \caption{Illustration of mapping alignments to attention masks for cross-attention of streaming chunkwise AED mode. The alignments are generated on-the-fly using the forward-backward variables obtained from the streaming HAT mode during training. Colored frames indicate the attended positions.
 }
 \label{fig:mask}
\end{figure}

\subsubsection{Streaming AED mode in All-in-One ASR}
\label{sssec:streaming_aed}
All-in-One ASR must support training in both offline and streaming modes, so we present here the training procedure for streaming chunkwise AED. 
As a modification, the encoder output sequence $\bm{H}^{\text{enc}}$ is replaced with its chunked variant $\bm{H}^{\text{enc}}_{i}$, as described  in \ref{ssec:encoder}, allowing the key and value matrices to be computed on a per-chunk basis. 
During training, the attention mask is applied to the cross-attention computation and is derived from the alignments, as illustrated in Fig.~\ref{fig:mask}. 

During decoding, our streaming AED can perform frame-synchronous decoding in the same manner as in the HAT mode by utilizing the blank probability shared with the HAT mode, without the need to recompute decoder states.
Specifically, this is achieved by simply replacing the HAT mode prediction $\hat{\bm{y}}_{t,u}^{\text{label (HAT)}}$ with the AED mode prediction $\hat{\bm{y}}_{u}^{\text{label (AED)}}$. 
Thus, our streaming AED mode does not require thresholding or triggers to determine the timing of label probability computation~\cite{chiu2018mocha,moritz19trigger,niko2019beamalgo,niko2020streaming,moriya2021rnnts2s,moriya2022rnntadlm}, nor does it depend on special tokens or explicit modeling to represent chunk or segment boundaries~\cite{Hsiao2020OnlineAS,Tsunoo2020StreamingTA,zeyer2022segmental_attention,zeineldeen2024chunked-aed}.
This efficiency is enabled by the fact that the multi-mode joiner is simply stacked on top of the encoder and predictor networks, allowing the predictor states to be fully shared across the HAT, AED, and LM modes.

\subsection{CTC mode}
\label{ssec:ctc_mode}
We employ a factored variant of CTC~\cite{hou2023CTCBlankThresholding,moriya2024hat_iam} for the CTC mode of the multi-mode joiner. The CTC mode also separately outputs blank probability $\hat{y}_{t}^{\text{blank (CTC)}} \in (0,1)^{1}$ and label probabilities $\hat{\bm{y}}_{t}^{\text{label (CTC)}} \in (0,1)^{K}$. The key difference from the HAT mode is that the transformed predictor output $\bm{h}^{\text{pred}^{\prime}}_{u}$ in Eq. (\ref{eq:advanced_pred_out}) is zeroed out. 
Thus, the sigmoid attention weight in Eq.~(\ref{eq:sigmoid_attention}) becomes a constant 0.5. 
The CTC mode in the multi-mode joiner is defined as follows:
\begin{eqnarray}
\bm{c}_{t}^{\text{Head}_h \text{(CTC)}} &=& 0.5 \cdot \bm{v}_{t}^{\text{Head}_h}, \\
\bm{c}_{t}^{\text{(CTC)}} &=& \bm{W}^{\text{proj}} \left[ \bm{c}_{t}^{\text{Head}_{1} \text{(CTC)} \top}, \cdots, \bm{c}_{t}^{\text{Head}_{d^{h}} \text{(CTC)} \top}\right]^{\top} , \\
\bm{h}_{t}^{\text{joiner (CTC)}} &=& \text{tanh} (\bm{c}_{t}^{\text{(CTC)}}),  \\
\bm{h}_{t}^{\text{joiner}^{\prime} \text{(CTC)}} &=&  \text{FeedForward} (\bm{h}_{t}^{\text{joiner (CTC)}} ; \theta^{\text{FF}}), \\
\bm{h}_{t}^{\text{joiner}^{\prime \prime} \text{(CTC)}} &=&  \text{LayerNorm} (\bm{h}_{t}^{\text{joiner (CTC)}} + \bm{h}_{t}^{\text{joiner}^{\prime} \text{(CTC)}} ; \theta^{\text{LN}_{\text{FF}}}), \\
\hat{y}_{t}^{\text{blank \ (CTC)}} &=& \text{Sigmoid} (\bm{w}^{\text{blank}} \bullet \bm{h}_{t}^{\text{joiner (CTC)}} + b^{\text{blank}} ), \\ 
\hat{\bm{y}}_{t}^{\text{label \ (CTC)}} &=& \text{Softmax} (\bm{W}^{\text{label}} \bm{h}_{t}^{\text{joiner}^{\prime \prime} \text{(CTC)}}  + \bm{b}^{\text{label}}), \label{eq:label_prob_ctc} \\
\hat{\bm{y}}^{\text{(CTC)}}_{t} &=& \left[ \ \hat{y}_{t}^{\text{blank \ (CTC)}}, \ (1 - \hat{y}_{t}^{\text{blank \ (CTC)}}) \cdot \hat{\bm{y}}_{t}^{\text{label \ (CTC)} \ \top} \ \right]^{\top}, \nonumber \\
\end{eqnarray}
where value vector $\bm{v}_{t}$ is the same as in Eq. (\ref{eq:advanced_value}).
The multi-mode joiner in the CTC mode outputs two dimensional matrix $\hat{\bm{Y}}^{\text{CTC}} \in (0,1)^{T \times (K+1)}$ during training. 
The parameter set $\Theta^{\text{HAT}^{\prime}}$ is optimized using CTC loss $\mathcal{L}_{\text{CTC}}$~\cite{Graves2006}. 

\subsection{LM mode}
\label{ssec:lm_mode}
The internal LM framework proposed for the HAT factorization~\cite{Ehsan2020hat} serves as the counterpart to the CTC mode, as described above. 
The LM mode, unlike the CTC mode, zeros out the key and value vectors that depend on the encoder output $\bm{h}^{\text{enc}^{\prime}}_{t}$ in Eq. (\ref{eq:advanced_enc_out}), and the prediction is given by $\hat{\bm{y}}_{u}^{\text{label (LM)}} \in (0,1)^{K}$, as follows:
\begin{eqnarray}
\bm{h}_{u}^{\text{joiner (LM)}} &=& \text{tanh} ( \bm{h}^{\text{pred}^{\prime}}_{u} ), \\
\bm{h}_{u}^{\text{joiner}^{\prime} \text{(LM)}} &=&  \text{FeedForward} (\bm{h}_{u}^{\text{joiner (LM)}} ; \theta^{\text{FF}}), \label{eq:ff_lm} \\
\bm{h}_{u}^{\text{joiner}^{\prime \prime} \text{(LM)}} &=&  \text{LayerNorm} (\bm{h}_{u}^{\text{joiner (LM)}} + \bm{h}_{u}^{\text{joiner}^{\prime} \text{(LM)}} ; \theta^{\text{LN}_{\text{FF}}}), \nonumber \\
\\
\hat{\bm{y}}_{u}^{\text{label \ (LM)}} &=& \text{Softmax} (\bm{W}^{\text{label}} \bm{h}_{u}^{\text{joiner}^{\prime \prime} \text{(LM)}}  + \bm{b}^{\text{label}}), \label{eq:label_prob_lm}
\end{eqnarray}
where $\bm{h}^{\text{pred}^{\prime}}_{u}$ is the same as in Eq. (\ref{eq:advanced_pred_out}). 
The predictions of the LM mode joiner form two-dimensional matrix $\hat{\bm{Y}}^{\text{LM}} \in (0,1)^{U \times K}$, and the parameter is optimized using cross-entropy loss $\mathcal{L}_{\text{LM}}$.

\subsection{Transducer with Attention (TwA) mode}
\label{ssec:twa_mode}
The proposed All-in-One ASR framework unifies the Transducer and AED models into a single model. 
Therefore, as demonstrated in prior work~\cite{prabhavalkar2017transducer_with_attention}, combining the Transducer with Attention (TwA) represents a natural and straightforward extension. 
Blank probability $\hat{y}_{t,u}^{\text{blank \ (TwA)}}$ and label probabilities $\hat{\bm{y}}_{t,u}^{\text{label \ (TwA)}}$ of the TwA model are defined as follows:
\begin{eqnarray}
\bm{h}_{t,u}^{\text{joiner (TwA)}} &=& \text{tanh} ( \bm{h}^{\text{enc}^{\prime}}_{t} + \bm{h}^{\text{pred}^{\prime}}_{u} + \bm{c}_{u}^{\text{(AED)}}) , \label{eq:joiner_space_twa} \\
\bm{h}_{t,u}^{\text{joiner}^{\prime} \text{(TwA)}} &=&  \text{FeedForward} (\bm{h}_{t,u}^{\text{joiner (TwA)}} ; \theta^{\text{FF}}), \\
\bm{h}_{t,u}^{\text{joiner}^{\prime \prime} \text{(TwA)}} &=&  \text{LayerNorm} (\bm{h}_{t,u}^{\text{joiner (TwA)}} + \bm{h}_{t,u}^{\text{joiner}^{\prime} \text{(TwA)}} ; \theta^{\text{LN}_{\text{FF}}}), \\
\hat{y}_{t,u}^{\text{blank \ (TwA)}} &=& \text{Sigmoid} (\bm{w}^{\text{blank}} \bullet \bm{h}_{t,u}^{\text{joiner (TwA)}} + b^{\text{blank}} ), \\ 
\hat{\bm{y}}_{t,u}^{\text{label \ (TwA)}} &=& \text{Softmax} (\bm{W}^{\text{label}} \bm{h}_{t,u}^{\text{joiner}^{\prime \prime} \text{(TwA)}}  + \bm{b}^{\text{label}}), \label{eq:label_prob_twa} \\
\hat{\bm{y}}^{\text{(TwA)}}_{t,u} &=& \left[ \ \hat{y}_{t,u}^{\text{blank \ (TwA)}}, \ (1 - \hat{y}_{t,u}^{\text{blank \ (TwA)}}) \cdot \hat{\bm{y}}_{t,u}^{\text{label \ (TwA)} \ \top} \ \right]^{\top}, \nonumber \\
\end{eqnarray}
where $\bm{h}^{\text{enc}^{\prime}}_{t}$, $\bm{h}^{\text{pred}^{\prime}}_{u}$, and $\bm{c}_{u}^{\text{(AED)}}$ are the same as in Eqs.~(\ref{eq:advanced_enc_out}), (\ref{eq:advanced_pred_out}), and (\ref{eq:context_vector_aed}), respectively. Compared to Eq. (\ref{eq:joiner_space_aed}), the difference from the AED mode is that $\bm{h}^{\text{enc}^{\prime}}_{t}$ is included inside $\tanh(\cdot)$ in Eq.~(\ref{eq:joiner_space_twa}).
Since the output tensor has the same dimensionality as in HAT, it can be optimized using the transducer loss $\mathcal{L}_{\text{TwA}}$. 
In conclusion, while TwA loss helps learn alignments in both the HAT and AED modes through joint training and enhances their ASR performance, its own ASR performance remains comparable to that of HAT and AED.\footnote{We evaluated the TwA mode within the All-in-One ASR models using the models marked with ``$\star$'' in Table~\ref{tab:summary_on_tlv2} and~\ref{tab:librispeech}, on the TLv2 and LibriSpeech test sets.
The WERs [\%] in offline and streaming modes were 6.9 and 7.8 for TLv2, and 2.5/5.4 and 2.8/6.8 for LibriSpeech, respectively.}  
Thus, our experiments do not focus on the ASR performance of TwA.

\subsection{Joint training for All-in-One ASR}
\label{ssec:jt}
Since most studies on joint training designate a primary decoder~\cite{watanabe2017hybrid,tara2019twopass,tanaka2019aje,moritz19trigger,niko2019beamalgo,niko2020streaming,moriya2021rnnts2s,moriya2022rnntadlm,moritz2023meta_rnnt,sudo2025asr4d}, the loss balancing weight assigned to it is typically larger than those of the other decoders.
Although the primary decoder tends to be stronger, our goal is to achieve a single unified model with high ASR performance across all ASR modes, in both offline and streaming modes.
Some studies in which multiple decoders are treated as primary have reported that this requires multi-stage training and careful tuning of the loss balancing weights~\cite{tara2019twopass,moriya2021rnnts2s,moriya2022rnntadlm,sudo2025asr4d}. 

In this work, we treat all decoders as primary ASR modes and investigate a joint training strategy that assigns equal weights to all ASR losses. 
Note that debiasing the weights of the subtask ASR decoder loss and balancing the weights across varying loss ranges for each ASR decoder~\cite{sudo2025asr4d} are left for future work. 
This training is performed in a single stage and is formulated as follows:
\begin{eqnarray}
\mathcal{L} &=& \mathcal{L}^{\text{Offline}} + \mathcal{L}^{\text{Streaming}} + \lambda \cdot \mathcal{L}_{\text{LM}}, \label{eq:total_loss_three} \\
\mathcal{L}^{*} &=&  \mathcal{L}^{*}_{\text{HAT}} + \mathcal{L}^{*}_{\text{AED}} + \mathcal{L}^{*}_{\text{CTC}} + \mathcal{L}^{*}_{\text{TwA}}, 
\label{eq:joint}
\end{eqnarray}
where ``*'' indicates either the offline or streaming mode.
Since it has been reported that increasing the internal LM factor can degrade ASR performance~\cite{meng2021ilmt,rui2023factorized_nt_adapt}, only the LM mode loss $\mathcal{L}_{\text{LM}}$ is weighted by coefficient $\lambda$, which is set to 0.1.

\subsection{Joint decoding for All-in-One ASR}
\label{ssec:joint_decoding}
In our All-in-One ASR framework, while decoding can be performed separately using each individual mode, joint decoding that leverages the strengths of all modes is also possible without recomputing the decoder states.
In this work, we apply shallow fusion~\cite{caglar2015shallow_fusion,kannan2018sfusion} with an external LM (ExtLM), incorporating LM subtraction~\cite{erik2019drf,Ehsan2020hat,meng2021ilme}.
For joint decoding, the $t$-th timestep and $u$-th label probabilities $\hat{\bm{y}}_{t,u}^{\text{label \ (HAT)}}$ in Eq. (\ref{eq:label_prob_hat}) are replaced with $\hat{\bm{y}}_{t,u}^{\text{label \ (HAT)}^{\prime}}$, as follows:
\begin{eqnarray}
  \hat{\bm{y}}_{t,u}^{\text{label (HAT)}^{\prime}} =    \hat{\bm{y}}_{t,u}^{\text{label (HAT)}^{\mu_{\text{HAT}}}} \odot \hat{\bm{y}}_{u}^{\text{label (AED)}^{\mu_{\text{AED}}}} \odot \frac{\hat{\bm{y}}_{u}^{\text{label (ExtLM)}^{\rho_{\text{ExtLM}}}}}{\hat{\bm{y}}_{u}^{\text{label (LM)}^{\rho_{\text{LM}}}}}, \label{eq:joint_decoding}
\end{eqnarray}
where $\hat{\bm{y}}_{t,u}^{\text{label (HAT)}}$, $\hat{\bm{y}}_{u}^{\text{label (AED)}}$, and $\hat{\bm{y}}_{u}^{\text{label (LM)}}$ denote the label probabilities in the HAT, AED, and LM modes, respectively, and are obtained from Eqs.~(\ref{eq:label_prob_hat}), (\ref{eq:label_prob_aed}), and (\ref{eq:label_prob_lm}). 
$\hat{\bm{y}}_{u}^{\text{label (ExtLM)}}$ represents the label probabilities from ExtLM. 
$\odot$ indicates the Hadamard product operation.
$\mu_{*}$ and $\rho_{*}$ are tunable hyperparameters, where $0 \leq \mu_{*} \leq 1$, $\sum \mu_{*} = 1$, and $\rho_{*} \geq 0$. 
Note that while the CTC objective was effective for training, incorporating the CTC prefix score for joint decoding did not lead to any improvement.


\begin{table}[t]
  \caption{Comparisons of Single-mode ASR (separate models for each ASR mode) and All-in-One ASR (a unified model for all ASR modes), and with/without Dual-mode support (unified offline and streaming modes), on the TLv2 test set. All models employed the multi-mode joiner architecture.}
  \vspace{-0.1cm}
  \label{tab:summary_on_tlv2}
  \centering
  \scalebox{0.88}[0.88]{
\hspace{-0.25cm} \begin{tabular}{@{}l|c@{\hspace{4pt}}c@{\hspace{4pt}}c|c@{\hspace{4pt}}c@{\hspace{4pt}}c|l@{}}
\hline
                           & \multicolumn{3}{c|}{Offline mode} & \multicolumn{3}{c|}{Streaming mode} & \multirow{2}{*}{\begin{tabular}[c]{@{}c@{}}$\sum$ of\\ \#param\end{tabular}} \\
Encoder \textbackslash \ ASR mode         & HAT       & AED       & CTC      & HAT        & AED       & CTC       &                                                                            \\ \hline
Conformer (Single)         & 6.9       & 7.1       & 7.6      & 8.7        & -         & 9.1       & 576.3M$^{\dagger}$                                                                       \\
ConMamba (Single)          & 6.9       & 7.1       & 7.7      & 8.5        & -         & 9.3       & 574.3M$^{\dagger}$                                                                     \\ \hline \hline
ConMamba (All-in-One w/o Dual)      & 6.9       & 6.9       & 7.6      & 8.1        & 8.2       & 9.2          & 235.8M$^{\ddagger}$                                                                     \\ \hline \hline
ConMamba (All-in-One)$^{\star}$ & \textbf{6.8}       & \textbf{6.8}       & \textbf{7.5}      & \textbf{7.8}        & \textbf{7.6}       & \textbf{8.9}      & 117.9M \\ \hline 
\multicolumn{8}{@{}l@{}}{$\dag$ and $\ddag$ indicate the sum of the parameters from five and two models, respectively.}
\end{tabular}
}
\vspace{-0.1cm}
\end{table}

\section{Experimental evaluations}
\label{sec:result}
\subsection{Data}
\label{ssec:data}
We evaluated our proposed All-in-One ASR frameworks on TED-LIUM release-2 (TLv2)~\cite{tedlium2} and LibriSpeech~\cite{librispeech} using ESPnet~\cite{espnet}.
The input feature was an 80-dimensional log Mel-filterbank extracted with a window size of 25ms at every 10ms, and augmentation methods~\cite{povey2015sp,specaugment,park2020adaptivespecaug} were applied during training. 
For tokenizing the output units, we adopted the word-piece model~\cite{sennrich2016bpe}, with vocabulary sizes of 500 and 1,000 for TLv2 and LibriSpeech, respectively. 

\subsection{System configuration}
\label{ssec:asrsystem}
For the ASR model, as illustrated in Fig.\ref{fig:architecture} (a) and detailed in~\ref{ssec:encoder}, we employed 17 self-attention-based Conformer blocks~\cite{anmol2020conformer} and 12 attention-free Conformer blocks~\cite{zhang2024conmamba,jiang2025speechslytherin_conmamba}, with the parameter sizes of both configurations matched to approximately 110M. 
As a common configuration, a two-layer 2-dimensional convolutional network with 256 filters was used, with a frame reduction rate of 4. 
The kernel size within the convolution module was set to 15. 
The model dimension and expansion factor within the feed-forward module were set at 512 and 4, respectively.
The Conformer includes 8 self-attention heads. 
For our ConMamba, the state space dimension, convolution dimension width, and expansion factor within each directional Mamba module were set to 16, 4, and 2, respectively. 
The predictor consisted of a two-layer 512-dimensional LSTM or Mamba, as shown in Fig.~\ref{fig:architecture} (b). 
The Mamba-based predictor contains root mean square normalization (RMSNorm)~\cite{biao2019rmsnorm}. 
In the streaming mode, chunk size was set to 20, corresponding to the frame length after convolutional subsampling, which resulted in an average algorithmic latency of 400ms. Note that the streaming Conformer utilized a history chunk of 20 as well.

All ASR models were initialized from scratch. The HAT, CTC, and LM modes were optimized using the pruned transducer~\cite{kuang2022pruned_rnnt}, internal acoustic and language modeling losses~\cite{Ehsan2020hat,meng2021ilmt,moriya2024hat_iam}, respectively. 
We used the Adam optimizer~\cite{Kingma2014adam} with a warm-up phase of 25k steps and a peak learning rate of 1.5e-3. Training was conducted for a total of 100 epochs on TLv2 and 200 epochs on LibriSpeech. 
We used alignment-length synchronous decoding~\cite{saon2020alsd,moriya2025lcssm_dual}, except for the offline AED mode, which employed label-synchronous decoding. A beam size of 8 was used in all cases.
We evaluated performance in terms of word error rate (WER). 

For the joint decoding, we also used on-the-fly internal LM estimation (ILME)~\cite{Ehsan2020hat,meng2021ilme}. 
The ExtLM consisted of 12 ConMamba encoder blocks. For the ConMamba-LM, the convolutional subsampling layers, BiMamba, and depthwise convolution within the ASR model's ConMamba blocks were simply replaced with an embedding layer, unidirectional Mamba, and causal depthwise convolution, respectively. 
The ExtLMs for each task were trained using the same datasets provided in the ESPnet recipes. 
The ASR and LM weights in Eq. (\ref{eq:joint_decoding}) were tuned on the development set.

 \vspace{-0.1cm}
\subsection{Results on TLv2} 

First, we present a summary of the results on the TLv2 test set in Table~\ref{tab:summary_on_tlv2}. 
All Single-mode ASR models were built separately, and the HAT and AED models were jointly trained with the CTC loss~\cite{watanabe2017hybrid,boyer2021rnntctc} for fair comparison. 
The number of cross-attention heads in the multi-mode joiner was set to 8 for all Single-mode ASR and All-in-One ASR models.
The attention-based Single-mode ASR (Conformer) was competitive with the attention-free counterpart using ConMamba. 
We also trained All-in-One ASR models without Dual-mode support using ConMamba, separately built for either offline or streaming mode. Their performance was competitive with that of the corresponding Single-mode ASR models. 
By incorporating Dual-mode support into the All-in-One ASR framework, the performance of all ASR modes in streaming mode was improved, as shown in~\cite{yu2021dualmode}, while maintaining the performance in offline mode. 
This results in reducing the model size of the All-in-One ASR to one-fifth that of the Single-mode ASR, which requires separately built models.

\begin{table}[t]
  \caption{Ablation study on the attention heads in multi-headed cross-attention within the multi-mode joiner. All models employed the ConMamba encoder and Mamba predictor.
  }
  \vspace{-0.1cm}
  \label{tab:ablation_attention_heads}
  \centering
  \scalebox{0.90}[0.90]{
\begin{tabular}{l|ccc|ccc}
\hline
        & \multicolumn{3}{c|}{Offline mode} & \multicolumn{3}{c}{Streaming mode} \\
\#heads & HAT       & AED       & CTC      & HAT        & AED       & CTC      \\ \hline
1       & 9.6       & 7.5       & 7.5      & 10.8       & 8.3       & 9.3      \\
4       & 6.8       & 6.9       & 7.4      & 7.8        & 7.7       & 8.8      \\
8       & 6.8       & 6.8       & 7.5      & 7.8        & 7.6       & 8.9      \\
16      & 6.7       & 7.1       & 7.5      & 7.8        & 7.8       & 9.0      \\ \hline
\end{tabular}
}
\vspace{-0.1cm}
\end{table}

We also investigated the effectiveness of varying the number of attention heads within the multi-mode joiner; the results are presented in Table~\ref{tab:ablation_attention_heads}. 
When the number of attention heads was set to 1, the performance of the HAT mode degraded significantly. This is because, under a single-head configuration, the sigmoid attention scores are insufficient to capture the complexity of multiple ASR modes, resulting in a tendency for the attention weights in Eq.~(\ref{eq:sigmoid_attention}) to collapse to zero. 
The models with 4 or more attention heads yielded approximately the same WERs. 
Hereafter, we employed the multi-mode joiner with 8 attention heads.

\begin{figure}[t]
    \hspace{-0.2cm} \includegraphics[width=10.8cm]{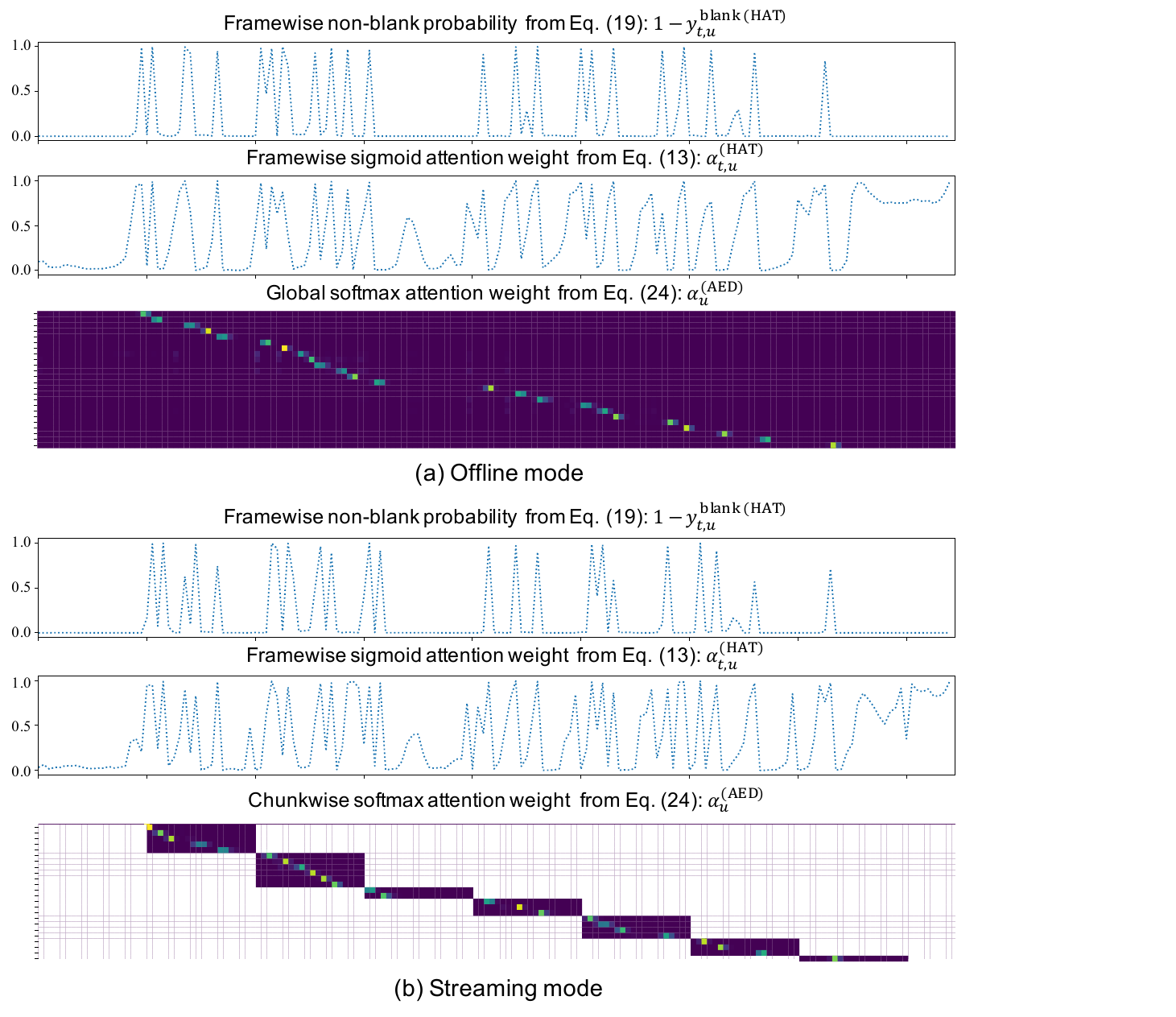}
\vspace{-0.6cm}
 \caption{Plots of the non-blank probabilities and sigmoid attention weights for the HAT mode, and the softmax attention weights for the AED mode, in (a) offline and (b) streaming modes, using the All-in-One ASR model marked with ``$\star$'', as presented in Table~\ref{tab:summary_on_tlv2}, generated from the utterance-ID ``AimeeMullins\_2009P-0124111-0124794'' (24 tokens) in the TLv2 test set.
 }
 \vspace{-0.2cm}
 \label{fig:attention}
\end{figure}

We conducted an analysis of each attention mechanism using the All-in-One ASR model marked with ``$\star$'', as presented in Table~\ref{tab:summary_on_tlv2}; the corresponding weights in greedy search decoding are plotted in Fig.~\ref{fig:attention}. 
We observe that the sigmoid attention weights exhibit behavior similar to the non-blank probabilities in the HAT mode. Additionally, the sigmoid attention weights in offline mode are more accurate, showing fewer redundant spikes compared to those in streaming mode. Furthermore, the attended timings of the sigmoid attention are slightly earlier than those of the softmax attention. We also find that the sigmoid attention weights increase near the end of speech, likely because the HAT model is trained with the explicit end-of-sentence token, and the sigmoid attention may be capturing the timing of decoding termination.

\begin{table}[t]
  \caption{Comparisons of encoder and predictor types, and joint decoding with/without ExtLM, in the All-in-One ASR model on the TLv2 test set. All models employed the multi-mode joiner.}
  \vspace{-0.1cm}
  \label{tab:comp_encoder_predictor}
  \centering
  \scalebox{0.88}[0.88]{
 \begin{tabular}{@{}l@{\hspace{2pt}}c@{\hspace{2pt}}c@{\hspace{2pt}}c@{\hspace{2pt}}|c@{\hspace{2pt}}c@{\hspace{2pt}}c|c@{\hspace{2pt}}c@{\hspace{2pt}}c@{}|@{}c@{}}
\hline
\multirow{2}{*}{Encoder} & \multirow{2}{*}{Predictor} & \multirow{2}{*}{\begin{tabular}[c]{@{}c@{}}Joint\\ Decode\end{tabular}} & \multirow{2}{*}{ExtLM} & \multicolumn{3}{c|}{Offline mode} & \multicolumn{3}{c|}{Streaming mode} & \multirow{2}{*}{\begin{tabular}[c]{@{}c@{}} $\sum$ of\\ \#param\end{tabular}} \\
                         &                            &                                                                           &                        & HAT       & AED       & CTC      & HAT        & AED       & CTC       &                                                                            \\ \hline
Conformer                & LSTM                       & -                                                                         & -                      & 6.9       & 7.6       & 7.6      & 8.1        & 8.2       & 9.0       & 118.9M                                                                           \\
Conformer                 & Mamba                      & -                                                                         & -                      & 6.9       & 7.4       & 7.7      & 8.1        & 8.1       & 9.1       & 118.3M                                                                     \\
ConMamba                & LSTM                       & -                                                                         & -                      & 7.0       & 7.0       & 7.6      & 7.9        & 7.9       & \textbf{8.8}       & 118.5M                                                                           \\
ConMamba                 & Mamba                      & -                                                                         & -                      & \textbf{6.8}       & \textbf{6.8}       & \textbf{7.5}      & \textbf{7.8}        & \textbf{7.6}       & 8.9       & 117.9M                                                                     \\ \hline \hline
ConMamba                 & Mamba                      & \checkmark                                                                 & -                      & 6.5 & - & -         & 7.3 & - & -          & 117.9M                                                                     \\ \hline \hline
ConMamba                 & Mamba                      & \checkmark                                                                 &  \checkmark                   & 5.9 & - & -        & 6.6 & - & -           & +80.9M                                                                           \\ \hline
\end{tabular}
}
\vspace{-0.1cm}
\end{table}

We conducted a comparison of the encoder and predictor types by using All-in-One ASR models with the multi-mode joiner, and the results are shown in Table~\ref{tab:comp_encoder_predictor}. 
Although the Single-mode ASR models with the Conformer encoder were competitive with those using the ConMamba encoder in Table~\ref{tab:summary_on_tlv2}, the ConMamba encoder within the All-in-One ASR models demonstrated superior capability compared to the Conformer encoder. 
This is likely because Mamba not only enables efficient modeling of long sequences, but also has the capability to simultaneously accommodate various types of modeling (i.e., HAT, AED, and CTC), each with distinct characteristics during training and decoding.\footnote{In the LibriSpeech experiment, the All-in-One ASR model with a Conformer encoder did not converge in training loss.} 
Switching from the LSTM to the Mamba predictor yielded slight improvements. 
Joint decoding using both HAT and AED modes further improved the WERs. 
Moreover, the ConMamba-LM provided additional gains, further reducing the WERs.

\begin{table}[t]
  \caption{WERs across utterance lengths for HAT mode, AED mode, and their joint decoding (HAT+AED) using the All-in-One ASR model without ExtLM in offline mode on the TLv2 test set.}
  \vspace{-0.1cm}
  \label{tab:results_tedlium_duration}
  \centering
  \scalebox{0.90}[0.90]{
\begin{tabular}{@{\hspace{2pt}}lc@{\hspace{8pt}}c@{\hspace{8pt}}c@{\hspace{8pt}}c@{\hspace{8pt}}c@{\hspace{8pt}}c|c@{\hspace{4pt}}}
\hline
\multirow{2}{*}{Mode} & \multicolumn{7}{c}{WERs {[}\%{]} for each duration range}                                                                                        \\ \cline{2-8} 
                      & -4s          & 4s-8s       & 8s-12s      & 12s-16s      & 16s-20s      & \multicolumn{1}{c|}{20s-}         & $\sum$          \\ \hline
HAT                   & 10.6         & 7.3          & 6.3          & 6.2          & 8.0          & 6.2          & 6.8          \\
AED                   & \textbf{9.9} & 7.1          & 6.4          & 5.9          & 8.5          & 8.4          & 6.8          \\
HAT+AED               & \textbf{9.9} & \textbf{7.0} & \textbf{6.2} & \textbf{5.6} & \textbf{7.5} & \textbf{5.9} & \textbf{6.5} \\ \hline
\end{tabular}
}
\vspace{-0.2cm}
\end{table}

Finally, we analyzed the effects of joint decoding using the All-in-One ASR model with the ConMamba encoder and Mamba predictor, as presented in Table~\ref{tab:comp_encoder_predictor}.
The WERs on the test set across different utterance duration ranges are shown in Table~\ref{tab:results_tedlium_duration}. 
We can see that the overall WERs for the offline HAT and AED modes were the same. However, although the All-in-One ASR uses identical parameters for decoding of each ASR mode, different error patterns were observed across various utterance duration ranges. 
The reason is that while HAT performs well for long utterances, AED is more effective for short utterances. 
In cases of unknown durations, those not seen in the training data (e.g., over 20s in the ESPnet TLv2 recipe), the global softmax cross-attention in the AED mode becomes unstable, leading to performance degradation, as noted in~\cite{chiu2019comp,inaguma2023AlignmentKD,stooke2024aligner}.
The joint decoding with the HAT and AED modes led to complementary improvements in WERs across most utterance duration ranges.

\subsection{Results on LibriSpeech}
\label{result_librispeech}

\begin{table}[t]
  \caption{Comparison of Single-mode ASR and All-in-One ASR (AiO ASR) on the LibriSpeech test sets (clean/other). All models employed the ConMamba encoder and the multi-mode joiner.}
  \vspace{-0.1cm}
  \label{tab:librispeech}
  \centering
  \scalebox{0.86}[0.86]{
\hspace{-0.3cm} \begin{tabular}{@{}l@{\hspace{2pt}}c@{\hspace{2pt}}c@{\hspace{2pt}}c@{\hspace{2pt}}|c@{\hspace{2pt}}c@{\hspace{4pt}}c|c@{\hspace{2pt}}c@{\hspace{4pt}}c@{\hspace{4pt}}|l@{}}
\hline
\multirow{2}{*}{Predictor} & \multirow{2}{*}{\begin{tabular}[c]{@{}c@{}}AiO\\ ASR\end{tabular}} & \multirow{2}{*}{\begin{tabular}[c]{@{}c@{}}Joint\\ Decode\end{tabular}} & \multirow{2}{*}{\begin{tabular}[c]{@{}c@{}}Ext\\ LM\end{tabular}} & \multicolumn{3}{c|}{Offline mode} & \multicolumn{3}{c|}{Streaming mode} & \multirow{2}{*}{\begin{tabular}[c]{@{}c@{}}$\sum$ of\\ \#param\end{tabular}} \\
                 &          &                                                                         &                        & HAT       & AED       & CTC      & HAT        & AED       & CTC       &                                                                            \\ \hline
Mamba      & -                & -                                                                       & -                      & \textbf{2.2}/\textbf{5.1}   & \textbf{2.4}/\textbf{5.2}   & 2.7/\textbf{6.0}  & 3.2/7.7    & -         & 3.9/8.6   & 576.4M$^{\dagger}$                                                                           \\ \hline \hline
LSTM   & \checkmark                    & -                                                                       & -                      & 2.3/5.5   & 2.6/5.6   & 2.7/6.4  & \textbf{2.8}/6.9    & 2.8/7.0   & 3.4/8.0          & 119.0M                                                                     \\
Mamba$^{\star}$  & \checkmark                    & -                                                                       & -                      & 2.3/5.2   & \textbf{2.4}/5.4   & \textbf{2.6}/6.1  & \textbf{2.8}/\textbf{6.7}    & \textbf{2.7}/\textbf{6.7}   & \textbf{3.2}/\textbf{7.8}   & 118.4M                                                                           \\ \hline \hline
Mamba  & \checkmark                    & \checkmark                                                               & -                      & 2.2/5.2 & - & -     & 2.7/6.5 & - & -      & 118.4M                                                                     \\ \hline \hline
Mamba  & \checkmark                    & \checkmark                                                               & \checkmark              & 2.0/4.7 & - & -     & 2.5/5.7 & - & -              & +81.4M                                                                     \\ \hline
\multicolumn{11}{@{}l@{}}{\ \dag \ The total consists of $118.4\text{M} \hspace{-2pt} \times \hspace{-2pt} 3 \ (\text{two HAT and one AED}) + 110.6\text{M} \hspace{-2pt} \times \hspace{-2pt} 2 \ (\text{two CTC})$.}
\end{tabular}
}
\vspace{-0.2cm}
\end{table}

We also evaluated our proposed methods on the LibriSpeech test sets; its results are shown in Table~\ref{tab:librispeech}. 
The ``AiO ASR'' column without and with a checkmark indicates the Single-mode ASR and All-in-One ASR models, respectively. 
In the LibriSpeech task, the Mamba predictor within the All-in-One ASR model also outperformed the LSTM predictor. 
As we can see, the All-in-One ASR model with the Mamba predictor achieved better streaming ASR performance than the Single-mode ASR models, while maintaining offline ASR performance without significant degradation.
Furthermore, joint decoding and ExtLM integration further reduced the WERs. 

\section{Conclusion and Discussions}
\label{ssec:conclusions}
We proposed the All-in-One ASR framework, which unifies various ASR modes in both offline and streaming modes into a single model by the introduction of a multi-mode joiner. 
Our framework simplifies model development and deployment, and significantly reduces the total number of parameters needed. 
The All-in-One ASR model achieved results that were competitive with or even better than those of individual single-mode ASR models in both offline and streaming modes.
Moreover, we confirmed that joint decoding is effective even in the All-in-One ASR model, even though it uses identical parameters across all ASR modes. 
We also examined ConMamba-LM, compatible with frame-synchronous decoding, and showed it further improved ASR performance.

In this work, we addressed the challenge of unifying diverse modeling characteristics, such as training objectives and decoding approaches, within a single model. 
Although this study focuses solely on ASR tasks, our ultimate goal is to extend this unification to other tasks within a single model to efficiently leverage shared and transferable knowledge. 
Future work includes the joint training of end-to-end tasks beyond single-talker ASR, such as Transducer-based applications~\cite{moriya2022tsrnnt,Kanda2022mtasr,moriya2025mtrnnt_aft,jiawei2021speecht,kanagawa2023vc_t,xue2022transducer_translation} and AED-based applications~\cite{delcroix2019tsasr_aed,masumura2023tsntsasr_aed,kanda2020sot_mtasr,wang2017first_aed_tts,e2e_translate,roshan2022speech_summarization,matsuura2024ssum}, within a single model using the All-in-One framework.

\newpage

\clearpage

\bibliographystyle{IEEEbib-abbrev}
\bibliography{refs}

@article{hochreiter1997lstm,
 author = {Hochreiter, Sepp and Schmidhuber, J\"{u}rgen},
 title = {{L}ong {S}hort-{T}erm {M}emory},
 journal = {Neural Computation},
 volume = {9},
 number = {8},
 year = {1997},
 pages = {1735--1780},
 publisher = {MIT Press},
 address = {Cambridge, MA, USA},
}

@inproceedings{Graves2006,
title = {{Connectionist Temporal Classification : Labelling Unsegmented Sequence Data with Recurrent Neural Networks}},
author = {Graves, Alex and Fernandez, Santiago and Gomez, Faustino and Schmidhuber, Jurgen},
booktitle = {Proc. of ICML},
pages = {369-376},
year = {2006},
memo = {CTC}
}

@InProceedings{Kingma2014adam,
author = {Kingma, Diederick P and Ba, Jimmy},
title = {{Adam: A method for stochastic optimization}},
booktitle = {Proc. of {ICLR}},
year = {2015}
}

@inproceedings{chorowski2014,
title={{End-to-End Continuous Speech Recognition Using Attention-based Recurrent NN: First Results}},
author={Chorowski, Jan and Bahdanau, Dzmitry and Cho, Kyunghyun and Bengio, Yoshua},
booktitle={Advances in {NIPS}},
year={2014}
}

@inproceedings{espnet,
  author    = {Shinji Watanabe and
               Takaaki Hori and
               Shigeki Karita and
               Tomoki Hayashi and
               Jiro Nishitoba and
               Yuya Unno and
               Nelson Enrique Yalta Soplin and
               Jahn Heymann and
               Matthew Wiesner and
               Nanxin Chen and
               Adithya Renduchintala and
               Tsubasa Ochiai},
  title     = {{ESP}net: End-to-End Speech Processing Toolkit},
  booktitle = {Proc. of {INTERSPEECH}},
  pages     = {2207--2211},
  year      = {2018}
}

@inproceedings{vaswani2017att,
  author    = {Ashish Vaswani and
               Noam Shazeer and
               Niki Parmar and
               Jakob Uszkoreit and
               Llion Jones and
               Aidan N. Gomez and
               Lukasz Kaiser and
               Illia Polosukhin},
  title     = {{Attention is All you Need}},
  booktitle = {Advances in {NIPS}},
  pages     = {5998--6008},
  year      = {2017}
}

@article{Kim2017JointCB,
  title={{Joint CTC-Attention based End-to-End Speech Recognition Using Multi-Task Learning}},
  author={Suyoun Kim and Takaaki Hori and Shinji Watanabe},
  journal={Proc. of {ICASSP}},
  year={2017},
  pages={4835-4839}
}

@inproceedings{Graves2012,
    Author = {Alex Graves}, 
    Booktitle = {Proc. of {ICML}},
    Title = {{Sequence Transduction with Recurrent Neural Networks}}, 
    Year = {2012}, 
}

@inproceedings{karita2020ctctransformer,
author = {Karita, Shigeki and Yalta, Nelson and Watanabe, Shinji and Delcroix, Marc and Ogawa, Atsunori and Nakatani, Tomohiro},
Booktitle={Proc. of {INTERSPEECH}},
year = {2019},
pages = {1408-1412},
title = {Improving Transformer-Based End-to-End Speech Recognition with Connectionist Temporal Classification and Language Model Integration}
}

@inproceedings{librispeech,
  title={{Librispeech: An {ASR} Corpus Based on Public Domain Audio Books}},
  author={Vassil Panayotov and Guoguo Chen and Daniel Povey and Sanjeev Khudanpur},
  booktitle={Proc. of {ICASSP}},
  year={2015},
  pages={5206-5210}
}

@article{watanabe2017hybrid,
title = {{Hybrid CTC/Attention Architecture for End-to-End Speech Recognition}},
author = "Shinji Watanabe and Takaaki Hori and Suyoun Kim and Hershey, {John R.} and Tomoki Hayashi",
year = "2017",
volume = "11",
pages = "1240--1253",
journal = "IEEE Journal on Selected Topics in Signal Processing",
number = "8"
}

@inproceedings{niko2020streaming,
author = {Moritz, Niko and Hori, Takaaki and Le Roux, Jonathan},
year = {2020},
title = {Streaming automatic speech recognition with the {T}ransformer model},
pages = {6069--6073},
booktitle = {Proc. of {ICASSP}}
}

@inproceedings{povey2015sp,
  author = {Ko, Tom and Peddinti, Vijayaditya and Povey, Daniel and Khudanpur, Sanjeev},
  booktitle = {Proc. of {INTERSPEECH}},
  pages = {3586-3589},
  title = {{Audio Augmentation for Speech Recognition}},
  year = 2015
}

@inproceedings{kannan2018sfusion,
author = {Kannan, Anjuli and Wu, Yonghui and Nguyen, Patrick and Sainath, Tara and Chen, ZhiJeng and Prabhavalkar, Rohit},
year = {2018},
pages = {5824-5828},
title = {An Analysis of Incorporating an External Language Model into a Sequence-to-Sequence Model},
booktitle = {Proc. of {ICASSP}}
}

@inproceedings{Ehsan2020hat, 
  author    = {Ehsan Variani and
               David Rybach and
               Cyril Allauzen and
               Michael Riley},
  title     = {{Hybrid Autoregressive {T}ransducer {(HAT)}}},
  booktitle = {Proc. of {ICASSP}},
  pages     = {6139--6143},
  year      = {2020}
}

@inproceedings{erik2019drf, 
  author    = {Erik McDermott and
               Hasim Sak and
               Ehsan Variani},
  title     = {A Density Ratio Approach to Language Model Fusion in End-to-End Automatic
               Speech Recognition},
  booktitle = {Proc. of {ASRU}},
  pages     = {434--441},
  year      = {2019}
}

@inproceedings{tedlium2,
    title = {{TED-LIUM: An Automatic Speech Recognition Dedicated Corpus}},
    author = "Rousseau, Anthony  and
      Del{\'e}glise, Paul  and
      Est{\`e}ve, Yannick",
    booktitle = "Proc. of {LREC}",
    year = "2012",
    pages = "125--129"
}

@inproceedings{yu2016lcblstm,
  author    = {Yu Zhang and
               Guoguo Chen and
               Dong Yu and
               Kaisheng Yao and
               Sanjeev Khudanpur and
               James R. Glass},
  title     = {Highway long short-term memory {RNN}s for distant speech recognition},
  booktitle = {Proc. of {ICASSP}},
  pages     = {5755--5759},
  year      = {2016}
}

@article{Xue2017lcblstm,
  title={Improving latency-controlled BLSTM acoustic models for online speech recognition},
  author={Shaofei Xue and Zhijie Yan},
  journal={Proc. of {ICASSP}},
  year={2017},
  pages={5340-5344}
}

@inproceedings{saon2020alsd,
author = {Saon, George and Tuske, Zoltan and Audhkhasi, Kartik},
year = {2020},
pages = {7799-7803},
title = {{Alignment-Length Synchronous Decoding for RNN Transducer}},
booktitle = {Proc. of {ICASSP}}
}

@inproceedings{tara2019twopass,
  author    = {Tara N. Sainath and
               Ruoming Pang and
               David Rybach and
               Yanzhang He and
               Rohit Prabhavalkar and
               Wei Li and
               Mirk{\'{o}} Visontai and
               Qiao Liang and
               Trevor Strohman and
               Yonghui Wu and
               Ian McGraw and
               Chung{-}Cheng Chiu},
  title     = {Two-Pass End-to-End Speech Recognition},
  booktitle = {Proc. of {INTERSPEECH}},
  pages     = {2773--2777},
  year      = {2019}
}

@inproceedings{tanaka2019aje,
  title={A Joint End-to-End and {DNN-HMM} Hybrid Automatic Speech Recognition System with Transferring Sharable Knowledge},
  author={T. Tanaka and Ryo Masumura and Takafumi Moriya and T. Oba and Y. Aono},
  booktitle={INTERSPEECH},
  pages={2210--2214},
  year={2019}
}

@inproceedings{chiu2019comp,
  author    = {Chung{-}Cheng Chiu and
               Anjuli Kannan and
               Rohit Prabhavalkar and
               Zhifeng Chen and
               Tara N. Sainath and
               Yonghui Wu and
               Wei Han and
               Yu Zhang and
               Ruoming Pang and
               Sergey Kishchenko and
               Patrick Nguyen and
               Arun Narayanan and
               Hank Liao and
               Shuyuan Zhang},
  title     = {A Comparison of End-to-End Models for Long-Form Speech Recognition},
  booktitle = {Proc. of {ASRU}},
  pages     = {889--896},
  year      = {2019}
}

@inproceedings{chiu2018mocha,
title={{Monotonic Chunkwise Attention}},
author={Chung-Cheng Chiu and Colin Raffel},
booktitle={Proc. of {ICLR}},
year={2018}
}

@inproceedings{moritz19trigger,
  author    = {Niko Moritz and
               Takaaki Hori and
               Jonathan Le Roux},
  title     = {Triggered Attention for End-to-end Speech Recognition},
  booktitle = {Proc. of {ICASSP}},
  pages     = {5666--5670},
  year      = {2019}
}

@inproceedings{niko2019beamalgo,
  author    = {Niko Moritz and
               Takaaki Hori and
               Jonathan Le Roux},
  title     = {Streaming End-to-End Speech Recognition with Joint {CTC}-{A}ttention Based Models},
  booktitle = {Proc. of {ASRU}},
  pages     = {936--943},
  year      = {2019}
}

@inproceedings{anmol2020conformer,
  author    = {Anmol Gulati and
               James Qin and
               Chung{-}Cheng Chiu and
               Niki Parmar and
               Yu Zhang and
               Jiahui Yu and
               Wei Han and
               Shibo Wang and
               Zhengdong Zhang and
               Yonghui Wu and
               Ruoming Pang},
  title     = {{Conformer: {C}onvolution-Augmented {T}ransformer for Speech Recognition}},
  booktitle = {Proc. of {INTERSPEECH}},
  pages     = {5036--5040},
  year      = {2020}
}

@inproceedings{moriya2022rnntadlm,
author = {Takafumi Moriya and Takanori Ashihara and Atsushi Ando and Hiroshi Sato and Tomohiro Tanaka and Kohei Matsuura and Ryo Masumura and Mark Delcroix and Takahiro Shinozaki},
title = {Hybrid {RNN-T}/{A}ttention-based streaming {ASR} with triggered chunkwise attention and dual internal language model integration},
Booktitle={Proc. of {ICASSP}},
pages={8282-8286},
year = {2022}
}

@inproceedings{moriya2021rnnts2s,
author = {Takafumi Moriya and Tomohiro Tanaka and Takanori Ashihara and Tsubasa Ochiai and Hiroshi Sato and Atsushi Ando and Ryo Masumura and Mark Delcroix and Taichi Asami},
title = {Streaming End-to-End Speech Recognition for {H}ybrid {RNN-T}/{A}ttention Architecture},
Booktitle={Proc. of {INTERSPEECH}},
pages={1787-1791},
year = {2021}
}

@inproceedings{meng2021ilmt,
  author    = {Zhong Meng and
               Naoyuki Kanda and
               Yashesh Gaur and
               Sarangarajan Parthasarathy and
               Eric Sun and
               Liang Lu and
               Xie Chen and
               Jinyu Li and
               Yifan Gong},
  title     = {{Internal Language Model Training for Domain-Adaptive End-To-End Speech
               Recognition}},
  booktitle = {Proc. of {ICASSP}},
  pages     = {7338--7342},
  year      = {2021}
}

@inproceedings{chen2021lcconformer,
  author    = {Xie Chen and
               Yu Wu and
               Zhenghao Wang and
               Shujie Liu and
               Jinyu Li},
  title     = {{Developing Real-Time Streaming {T}ransformer {T}ransducer for Speech Recognition on Large-Scale Dataset}},
  booktitle = {Proc. of {ICASSP}},
  pages     = {5904--5908},
  year      = {2021}
}

@inproceedings{meng2021ilme, 
  author    = {Zhong Meng and
               Sarangarajan Parthasarathy and
               Eric Sun and
               Yashesh Gaur and
               Naoyuki Kanda and
               Liang Lu and
               Xie Chen and
               Rui Zhao and
               Jinyu Li and
               Yifan Gong},
  title     = {{Internal Language Model Estimation for Domain-Adaptive End-to-End
               Speech Recognition}},
  booktitle = {Proc. of {SLT}},
  pages     = {243--250},
  year      = {2021}
}

@inproceedings{Sainath2021AnES,
  title={{An Efficient Streaming Non-Recurrent On-Device End-to-End Model with Improvements to Rare-Word Modeling}},
  author={Tara N. Sainath and Yanzhang He and Arun Narayanan and Rami Botros and Ruoming Pang and David Rybach and Cyril Allauzen and Ehsan Variani and James Qin and Quoc-Nam Le-The and Shuo-yiin Chang and Bo Li and Anmol Gulati and Jiahui Yu and Chung-Cheng Chiu and Diamantino Caseiro and Wei Li and Qiao Liang and Pat Rondon},
  booktitle={Proc. of {INTERSPEECH}},
  pages={1777--1781},
  year={2021}
}

@inproceedings{Kanda2022mtasr,
  author    = {Naoyuki Kanda and
               Jian Wu and
               Yu Wu and
               Xiong Xiao and
               Zhong Meng and
               Xiaofei Wang and
               Yashesh Gaur and
               Zhuo Chen and
               Jinyu Li and
               Takuya Yoshioka},
  title     = {{Streaming Multi-Talker {ASR} with Token-Level Serialized Output Training}},
  pages        = {3774--3778},
  booktitle = {Proc. of {INTERSPEECH}},
  year         = {2022}
}

@inproceedings{sennrich2016bpe,
    title = "{Neural Machine Translation of Rare Words with Subword Units}",
    author = "Sennrich, Rico  and
      Haddow, Barry  and
      Birch, Alexandra",
    booktitle = "Proc. of {ACL}",
    year = "2016",
    pages = "1715--1725"
}

@inproceedings{le2023BlankThresholding,
  author       = {Duc Le and
                  Frank Seide and
                  Yuhao Wang and
                  Yang Li and
                  Kjell Schubert and
                  Ozlem Kalinli and
                  Michael L. Seltzer},
  title        = {{Factorized Blank Thresholding for Improved Runtime Efficiency of Neural Transducers}},
  booktitle    = {Proc. of {ICASSP}},
  pages        = {1--5},
  year         = {2023}
}

@inproceedings{hou2023CTCBlankThresholding,
  author       = {Junfeng Hou and
                  Peiyao Wang and
                  Jincheng Zhang and
                  Meng Yang and
                  Minwei Feng and
                  Jingcheng Yin},
  title        = {{{CTC} Blank Triggered Dynamic Layer-Skipping for Efficient CTC-based
                  Speech Recognition}},
  booktitle    = {Proc. of {ASRU}},
  pages       = {1--5},
  year         = {2023}
}

@inproceedings{boyer2021rnntctc,
  author       = {Florian Boyer and
                  Yusuke Shinohara and
                  Takaaki Ishii and
                  Hirofumi Inaguma and
                  Shinji Watanabe},
  title        = {{A Study of Transducer Based End-to-End {ASR} with ESPnet: Architecture,
                  Auxiliary Loss and Decoding Strategies}},
  booktitle    = {Proc. of {ASRU}},
  pages        = {16--23},
  year         = {2021}
}

@inproceedings{moritz2023meta_rnnt,
  author       = {Niko Moritz and
                  Frank Seide and
                  Duc Le and
                  Jay Mahadeokar and
                  Christian Fuegen},
  title        = {{An Investigation of Monotonic Transducers for Large-Scale Automatic
                  Speech Recognition}},
  booktitle    = {Proc. of {SLT}},
  pages        = {324--330},
  year         = {2022}
}

@inproceedings{specaugment,
  title={{Spec{A}ugment: A Simple Data Augmentation Method for Automatic Speech Recognition}},
  author={Park, Daniel S and Chan, William and Zhang, Yu and Chiu, Chung-Cheng and Zoph, Barret and Cubuk, Ekin D and Le, Quoc V},
  journal={Proc. of {INTERSPEECH}},
  pages={2613--2617},
  year={2019}
}

@ARTICLE{rohit2024suvey,
  author={Prabhavalkar, Rohit and Hori, Takaaki and Sainath, Tara N. and Schlüter, Ralf and Watanabe, Shinji},
  journal={{IEEE/ACM Transactions on Audio, Speech, and Language Processing}}, 
  title={{End-to-End Speech Recognition: A Survey}}, 
  year={2024},
  volume={32},
  number={},
  pages={325-351}
}

@article{jinyu2022survey,
year = {2022},
volume = {11},
journal = {{APSIPA Transactions on Signal and Information Processing}},
title = {{Recent Advances in End-to-End Automatic Speech Recognition}},
number = {1},
author = {Jinyu Li}
}

@INPROCEEDINGS{rui2023factorized_nt_adapt,
  author={Zhao, Rui and Xue, Jian and Parthasarathy, Partha and Miljanic, Veljko and Li, Jinyu},
  booktitle={Proc. of {ICASSP}},
  title={{Fast and Accurate Factorized Neural Transducer for Text Adaption of End-to-End Speech Recognition Models}}, 
  year={2023},
  pages={1-5}
}

@article{inaguma2023AlignmentKD,
  author       = {Hirofumi Inaguma and
                  Tatsuya Kawahara},
  title        = {{Alignment Knowledge Distillation for Online Streaming Attention-Based
                  Speech Recognition}},
  journal      = {{IEEE} {ACM} Trans. Audio Speech Lang. Process.},
  volume       = {31},
  pages        = {1371--1385},
  year         = {2023}
}

@inproceedings{yu2021dualmode,
title={{Dual-mode ASR: Unify and Improve Streaming ASR with Full-context Modeling}},
author={Jiahui Yu and Wei Han and Anmol Gulati and Chung-Cheng Chiu and Bo Li and Tara N Sainath and Yonghui Wu and Ruoming Pang},
booktitle={Proc. of {ICML}}, 
year={2021}
}

@INPROCEEDINGS{yangyang2022emformer,
  author={Shi, Yangyang and Wu, Chunyang and Wang, Dilin and Xiao, Alex and Mahadeokar, Jay and Zhang, Xiaohui and Liu, Chunxi and Li, Ke and Shangguan, Yuan and Nagaraja, Varun and Kalinli, Ozlem and Seltzer, Mike},
  booktitle={Proc. of {ICASSP}},
  title={{Streaming Transformer Transducer based Speech Recognition Using Non-Causal Convolution}}, 
  year={2022},
  pages={8277-8281}
}

@INPROCEEDINGS{qian2020streaming_tt,
  author={Zhang, Qian and Lu, Han and Sak, Hasim and Tripathi, Anshuman and McDermott, Erik and Koo, Stephen and Kumar, Shankar},
  booktitle={Proc. of {ICASSP}},
  title={{Transformer Transducer: A Streamable Speech Recognition Model with Transformer Encoders and RNN-T Loss}}, 
  year={2020},
  pages={7829-7833}
}

@INPROCEEDINGS{strimel2023lookahead,
  title = 	 {{Lookahead When It Matters: Adaptive Non-causal Transformers for Streaming Neural Transducers}},
  author =       {Strimel, Grant and Xie, Yi and King, Brian John and Radfar, Martin and Rastrow, Ariya and Mouchtaris, Athanasios},
  booktitle = 	 {Proc. of {ICML}},
  pages = 	 {32654--32676},
  year = 	 {2023}
}

@inproceedings{jiang2025speechslytherin_conmamba,
      title={{Speech Slytherin: Examining the Performance and Efficiency of Mamba for Speech Separation, Recognition, and Synthesis}}, 
      author={Xilin Jiang and Yinghao Aaron Li and Adrian Nicolas Florea and Cong Han and Nima Mesgarani},
      year={2025},
      booktitle={Proc. of {ICASSP}},
      pages={1--5}
}

@article{zhang2024conmamba,
  title={{Mamba in Speech: Towards an Alternative to Self-Attention}},
  author={Zhang, Xiangyu and Zhang, Qiquan and Liu, Hexin and Xiao, Tianyi and Qian, Xinyuan and Ahmed, Beena and Ambikairajah, Eliathamby and Li, Haizhou and Epps, Julien},
  journal={arXiv preprint arXiv:2405.12609},
  year={2024}
}

@article{mamba,
  title={{Mamba: Linear-Time Sequence Modeling with Selective State Spaces}},
  author={Gu, Albert and Dao, Tri},
  journal={arXiv preprint arXiv:2312.00752},
  year={2023}
}

@inproceedings{shashi2025dualmode_xlsr,
author = {Kumar, Shashi and Madikeri, Srikanth and Zuluaga, Juan Pablo and Villatoro-Tello, Esaú and Nigmatulina, Iuliia and Motlicek, Petr and E, Manjunath and Ganapathiraju, Aravind},
year = {2025},
title = {{XLSR-Transducer: Streaming ASR for Self-Supervised Pretrained Models}},
booktitle = {Proc. of {ICASSP}},
pages     = {1--5}
}

@inproceedings{kuang2022pruned_rnnt,
  title     = {{Pruned RNN-T for fast, memory-eﬀicient ASR training}},
  author    = {Fangjun Kuang and Liyong Guo and Wei Kang and Long Lin and Mingshuang Luo and Zengwei Yao and Daniel Povey},
  year      = {2022},
  booktitle = {Proc. of {INTERSPEECH}},
  pages     = {2068--2072}
}

@inproceedings{biao2019rmsnorm,
 author = {Zhang, Biao and Sennrich, Rico},
 booktitle = {Advances in {NeurIPS}},
 title = {{Root Mean Square Layer Normalization}},
 year = {2019}
}

@inproceedings{moriya2024hat_iam,
  title     = {{Boosting Hybrid Autoregressive Transducer-based ASR with Internal Acoustic Model Training and Dual Blank Thresholding}},
  author    = {Takafumi Moriya and Takanori Ashihara and Masato Mimura and Hiroshi Sato and Kohei Matsuura and Ryo Masumura and Taichi Asami},
  year      = {2024},
  booktitle = {Proc. of {INTERSPEECH}},
  pages     = {3465--3469}
}

@article{Li2023DynamicConv,
  title={{Dynamic Chunk Convolution for Unified Streaming and Non-Streaming Conformer ASR}},
  author={Xilai Li and Goeric Huybrechts and S. Ronanki and Jeffrey J. Farris and S. Bodapati},
  journal={Proc. of {ICASSP}},
  year={2023},
  pages={1-5}
}

@INPROCEEDINGS{mimura2025attn_mamba,
  author={Mimura, Masato and Moriya, Takafumi and Matsuura, Kohei},
  booktitle={Proc. of {ICASSP}},
  title={{Advancing Streaming ASR with Chunk-wise Attention and Trans-chunk Selective State Spaces}},
  year={2025},
  pages={1-5}
}

@ARTICLE{sudo2025asr4d,
  author={Sudo, Yui and Shakeel, Muhammad and Fukumoto, Yosuke and Yan, Brian and Shi, Jiatong and Peng, Yifan and Watanabe, Shinji},
  journal={{IEEE/ACM Transactions on Audio, Speech, and Language Processing}},
  title={{Joint Beam Search Integrating CTC, Attention, and Transducer Decoders}}, 
  year={2025},
  volume={33},
  pages={598-612}
}

@inproceedings{prabhavalkar2017transducer_with_attention,
  title     = {{A Comparison of Sequence-to-Sequence Models for Speech Recognition}},
  author    = {Rohit Prabhavalkar and Kanishka Rao and Tara N. Sainath and Bo Li and Leif Johnson and Navdeep Jaitly},
  year      = {2017},
  booktitle = {Proc. of {INTERSPEECH}},
  pages     = {939--943}
}

@inproceedings{ramapuram2025sigmoid_attention,
title={{Theory, Analysis, and Best Practices for Sigmoid Self-Attention}},
author={Jason Ramapuram and Federico Danieli and Eeshan Gunesh Dhekane and Floris Weers and Dan Busbridge and Pierre Ablin and Tatiana Likhomanenko and Jagrit Digani and Zijin Gu and Amitis Shidani and Russell Webb},
booktitle={Proc. of {ICLR}},
year={2025}
}

@inproceedings{ioffe2015batchnorm,
  author       = {Sergey Ioffe and
                  Christian Szegedy},
  title        = {{Batch Normalization: Accelerating Deep Network Training by Reducing Internal Covariate Shift}},
  booktitle    = {Proc. of {ICML}},
  pages        = {448--456},
  year         = {2015}
}

@article{jimmy2016layernorm,
  author       = {Lei Jimmy Ba and
                  Jamie Ryan Kiros and
                  Geoffrey E. Hinton},
  title        = {{Layer Normalization}},
  journal      = {CoRR},
  volume       = {abs/1607.06450},
  year         = {2016},
  eprinttype    = {arXiv},
  eprint       = {1607.06450}
}

@inproceedings{saon2021mijoiner,
  title={{Advancing RNN Transducer Technology for Speech Recognition}},
  author={George Saon and Zoltan Tueske and Daniel Bola{\~n}os and Brian Kingsbury},
  booktitle={Proc. of {ICASSP}},
  year={2021},
  pages={5654-5658}
}

@inproceedings{stooke2024aligner,
  title={{Aligner-Encoders: Self-Attention Transformers Can Be Self-Transducers}},
  author={Adam Stooke and Rohit Prabhavalkar and Khe Chai Sim and Pedro J Moreno Mengibar},
  booktitle={Advances in {NeurIPS}},
  year={2024}
}

@InProceedings {zeineldeen2024chunked-aed,
author= {Zeineldeen, Mohammad and Zeyer, Albert and Schlüter, Ralf and Ney, Hermann}, 
title= {{Chunked Attention-based Encoder-Decoder Model for Streaming Speech Recognition}}, 
booktitle= {Proc. of {ICASSP}},
year= {2024}, 
pages={11331-11335}
}

@inproceedings{tang2023hybrid_rnnt_aed,
    title = {{Hybrid Transducer and Attention based Encoder-Decoder Modeling for Speech-to-Text Tasks}},
    author = {Tang, Yun  and
      Sun, Anna  and
      Inaguma, Hirofumi  and
      Chen, Xinyue  and
      Dong, Ning  and
      Ma, Xutai  and
      Tomasello, Paden  and
      Pino, Juan},
    booktitle = {Proc. of {ACL}},
    year = {2023},
    address = "Toronto, Canada",
    pages = {12441--12455}
}

@inproceedings{Hu2020Deliberation,
  title={{Deliberation Model Based Two-Pass End-To-End Speech Recognition}},
  author={Ke Hu and Tara N. Sainath and Ruoming Pang and Rohit Prabhavalkar},
  booktitle={Proc. of {ICASSP}},
  year={2020},
  pages={7799-7803}
}

@INPROCEEDINGS{park2020adaptivespecaug,
  author={Park, Daniel S. and Zhang, Yu and Chiu, Chung-Cheng and Chen, Youzheng and Li, Bo and Chan, William and Le, Quoc V. and Wu, Yonghui},
  booktitle={Proc. of {ICASSP}},
  title={Specaugment on Large Scale Datasets}, 
  year={2020},
  pages={6879-6883}
}

@inproceedings{tanaka2019joint_dnn_aed,
  title     = {{A Joint End-to-End and DNN-HMM Hybrid Automatic Speech Recognition System with Transferring Sharable Knowledge}},
  author    = {Tomohiro Tanaka and Ryo Masumura and Takafumi Moriya and Takanobu Oba and Yushi Aono},
  year      = {2019},
  booktitle = {Proc. of {INTERSPEECH}},
  pages     = {2210--2214},
  issn      = {2958-1796}
}

@article{Hsiao2020OnlineAS,
  title={Online Automatic Speech Recognition With Listen, Attend and Spell Model},
  author={Roger Hsiao and Dogan Can and Tim Ng and Ruchir Travadi and Arnab Ghoshal},
  journal={IEEE Signal Processing Letters},
  year={2020},
  volume={27},
  pages={1889-1893}
}

@inproceedings{Tsunoo2020StreamingTA,
  title={{Streaming Transformer ASR With Blockwise Synchronous Beam Search}},
  author={Emiru Tsunoo and Yosuke Kashiwagi and Shinji Watanabe},
  booktitle={Proc. of {SLT}},
  year={2020},
  pages={22-29}
}

@inproceedings{zeyer2022segmental_attention,
author= {Zeyer, Albert and Schmitt, Robin and Zhou, Wei and Schl\"uter, Ralf and Ney, Hermann}, 
title= {{Monotonic Segmental Attention for Automatic Speech Recognition}}, 
booktitle= {Proc. of {SLT}},
year= 2023, 
pages= {229-236}
}

@inproceedings{jiawei2021speecht,
 author = {Chen, Jiawei and Tan, Xu and Leng, Yichong and Xu, Jin and Wen, Guihua and Qin, Tao and Liu, Tie-Yan},
 booktitle = {Advances in {NeurIPS}},
 pages = {6621--6633},
 title = {{Speech-T: Transducer for Text to Speech and Beyond}},
 year = {2021}
}

@inproceedings{kanagawa2023vc_t,
  title     = {{VC-T: Streaming Voice Conversion Based on Neural Transducer}},
  author    = {Hiroki Kanagawa and Takafumi Moriya and Yusuke Ijima},
  year      = {2023},
  booktitle = {Proc. of {INTERSPEECH}},
  pages     = {2088--2092}
}

@inproceedings{e2e_translate,
  author       = {Alexandre Berard and
                  Olivier Pietquin and
                  Christophe Servan and
                  Laurent Besacier},
  title        = {{Listen and Translate: A Proof of Concept for End-to-End Speech-to-Text Translation}},
  booktitle      = {Proc. of {NeurIPS} Workshop on end-to-end learning for speech and audio processing},
  year         = {2016}
}

@INPROCEEDINGS{roshan2022speech_summarization,
  author={Sharma, Roshan and Palaskar, Shruti and Black, Alan W and Metze, Florian},
  booktitle={Proc. of {ICASSP}},
  title={{End-to-End Speech Summarization Using Restricted Self-Attention}}, 
  year={2022},
  pages={8072-8076}
}

@inproceedings{matsuura2024ssum,
  title     = {{Sentence-wise Speech Summarization: Task, Datasets, and End-to-End Modeling with LM Knowledge Distillation}},
  author    = {Kohei Matsuura and Takanori Ashihara and Takafumi Moriya and Masato Mimura and Takatomo Kano and Atsunori Ogawa and Marc Delcroix},
  year      = {2024},
  booktitle = {Proc. of {INTERSPEECH}},
  pages     = {1945--1949}
}

@inproceedings{xue2022transducer_translation,
  title     = {{Large-Scale Streaming End-to-End Speech Translation with Neural Transducers}},
  author    = {Jian Xue and Peidong Wang and Jinyu Li and Matt Post and Yashesh Gaur},
  year      = {2022},
  booktitle = {Proc. of {INTERSPEECH}},
  pages     = {3263--3267}
}

@article{caglar2015shallow_fusion,
author = {Gulcehre, Caglar and Firat, Orhan and Xu, Kelvin and Cho, Kyunghyun and Barrault, Loïc and Lin, Huei-Chi and Bougares, Fethi and Schwenk, Holger and Bengio, Y.},
year = {2015},
volume = {45},
journal = {Computer Speech \& Language},
pages = {137--148},
title = {{On Using Monolingual Corpora in Neural Machine Translation}}
}

@inproceedings{yuhuai2016mulitiplicative_integration,
 author = {Wu, Yuhuai and Zhang, Saizheng and Zhang, Ying and Bengio, Yoshua and Salakhutdinov, Russ R},
 booktitle = {Advances in {NeurIPS}},
 title = {{On Multiplicative Integration with Recurrent Neural Networks}},
 year = {2016}
}

@inproceedings{moriya2022tsrnnt,
  title     = {{Streaming Target-Speaker ASR with Neural Transducer}},
  author    = {Takafumi Moriya and Hiroshi Sato and Tsubasa Ochiai and Marc Delcroix and Takahiro Shinozaki},
  year      = {2022},
  booktitle = {Proc. of {INTERSPEECH}},
  pages     = {2673--2677}
}

@INPROCEEDINGS{moriya2025mtrnnt_aft,
  author={Moriya, Takafumi and Horiguchi, Shota and Delcroix, Marc and Masumura, Ryo and Ashihara, Takanori and Sato, Hiroshi and Matsuura, Kohei and Mimura, Masato},
  booktitle={Proc. of {ICASSP}},
  title={{Alignment-Free Training for Transducer-based Multi-Talker ASR}}, 
  year={2025},
  pages={1-5}
}

@inproceedings{kanda2020sot_mtasr,
  title     = {{Serialized Output Training for End-to-End Overlapped Speech Recognition}},
  author    = {Naoyuki Kanda and Yashesh Gaur and Xiaofei Wang and Zhong Meng and Takuya Yoshioka},
  year      = {2020},
  booktitle = {Proc. of {INTERSPEECH}},
  pages     = {2797--2801}
}

@inproceedings{delcroix2019tsasr_aed,
  title     = {{End-to-End SpeakerBeam for Single Channel Target Speech Recognition}},
  author    = {Marc Delcroix and Shinji Watanabe and Tsubasa Ochiai and Keisuke Kinoshita and Shigeki Karita and Atsunori Ogawa and Tomohiro Nakatani},
  year      = {2019},
  booktitle = {Proc. of {INTERSPEECH}},
  pages     = {451--455}
}

@inproceedings{masumura2023tsntsasr_aed,
  title     = {{End-to-End Joint Target and Non-Target Speakers ASR}},
  author    = {Ryo Masumura and Naoki Makishima and Taiga Yamane and Yoshihiko Yamazaki and Saki Mizuno and Mana Ihori and Mihiro Uchida and Keita Suzuki and Hiroshi Sato and Tomohiro Tanaka and Akihiko Takashima and Satoshi Suzuki and Takafumi Moriya and Nobukatsu Hojo and Atsushi Ando},
  year      = {2023},
  booktitle = {Proc.of {INTERSPEECH}},
  pages     = {2903--2907}
}

@inproceedings{wang2017first_aed_tts,
  title     = {{Tacotron: Towards End-to-End Speech Synthesis}},
  author    = {Yuxuan Wang and R.J. Skerry-Ryan and Daisy Stanton and Yonghui Wu and Ron J. Weiss and Navdeep Jaitly and Zongheng Yang and Ying Xiao and Zhifeng Chen and Samy Bengio and Quoc Le and Yannis Agiomyrgiannakis and Rob Clark and Rif A. Saurous},
  year      = {2017},
  booktitle = {Proc. of {INTERSPEECH}},
  pages     = {4006--4010}
}

@INPROCEEDINGS{li2021causal_online_model,
  author={Li, Bo and Gulati, Anmol and Yu, Jiahui and Sainath, Tara N. and Chiu, Chung-Cheng and Narayanan, Arun and Chang, Shuo-Yiin and Pang, Ruoming and He, Yanzhang and Qin, James and Han, Wei and Liang, Qiao and Zhang, Yu and Strohman, Trevor and Wu, Yonghui},
  booktitle={Proc. of {ICASSP}},
  title={{A Better and Faster end-to-end Model for Streaming ASR}}, 
  year={2021},
  pages={5634-5638}
}

@inproceedings{moriya2025lcssm_dual,
  title     = {{Attention-Free Dual-Mode ASR with Latency-Controlled Selective State Spaces}},
  author    = {Takafumi Moriya and Masato Mimura and Kiyoaki Matsui and Hiroshi Sato and Kohei Matsuura},
  year      = {2025},
  booktitle = {Proc.of {INTERSPEECH}},
  pages     = {3588--3592}
}

\end{document}